\documentclass[journal]{journal}
\pdfoutput=1
\usepackage{multicol,lipsum}
\ifCLASSINFOpdf
	\usepackage[pdftex]{graphicx}
 
\else
\fi
\usepackage[cmex10]{amsmath}

\usepackage{mathtools}

\usepackage{float}

\usepackage[export]{adjustbox}

\usepackage{fixltx2e}
\usepackage{ulem}
\usepackage{subfigure}
\usepackage{xcolor}
\usepackage{setspace}

\usepackage[colorlinks,
            linkcolor=blue,
            anchorcolor=blue,
            citecolor=blue]{hyperref}

\usepackage{nameref}

\begin{document}
%
\title{Newtonian Mechanics Based Transient Stability PART I: Machine Paradigms}
%
%
%

\author{{Songyan Wang,
        Jilai Yu,  
				Aoife Foley
        Jingrui Zhang
        }
        
}
%
%

\markboth{Journal of \LaTeX\ Class Files,~Vol.~6, No.~1, January~2007}%
{Shell \MakeLowercase{\textit{et al.}}: Bare Demo of IEEEtran.cls for Journals}
%



\maketitle
\thispagestyle{empty}
\begin{abstract}

Individual-machine, superimposed-machine and equivalent-machine can be seen as the three major perspectives of the power system transient stability.
In this paper, the machine paradigms are established according to the common thinking among the three different machines.
The machine paradigms comprise of the three components, i.e., trajectory paradigm, modeling paradigm and energy paradigm.
The trajectory paradigm is the reflection of the trajectory stability; the modeling paradigm is the two-machine-system modeling of the trajectory stability; and the energy paradigm is the stability evaluation of the two-machine system.
Based on this, it is clarified that the machine paradigms can be expressed into the individual machine form or the equivalent machine form. Then, the relationship between the machine stability and the system stability are analyzed.
Simulation results show that the effectiveness of both the individual-machine and the equivalent machine is fully based on the strict followings of the machine paradigms.
\end{abstract}

%

\IEEEpeerreviewmaketitle

  \begin{tabular}{lllll}
    &            &               &                  &                                \\
  \multicolumn{5}{c}{\textbf{Nomenclature}}                                                \\
  KE     & \multicolumn{1}{c}{}  & \multicolumn{3}{l}{Kinetic energy}                      \\
  PE     & \multicolumn{1}{c}{}  & \multicolumn{3}{l}{Potential energy}                    \\
  COI    & \multicolumn{1}{c}{}  & \multicolumn{3}{l}{Center of inertia}                   \\
  DLP    &                       & \multicolumn{3}{l}{Dynamic liberation point}            \\
  DSP    &                       & \multicolumn{3}{l}{Dynamic stationary point}            \\
  EAC    &                       & \multicolumn{3}{l}{Equal area criterion}                \\
  MPP    &                       & \multicolumn{3}{l}{Maximum potential energy point}      \\
  NEC    &                       & \multicolumn{3}{l}{Newtonian energy conversion}         \\
  TSA    &                       & \multicolumn{3}{l}{Transient stability assessment}      \\
  TEF    &                       & \multicolumn{3}{l}{Transient energy function}           \\
  SMKE   &                       & \multicolumn{3}{l}{Superimposed-machine kinetic energy} \\
  SMTE   &                       & \multicolumn{3}{l}{Superimposed-machine total energy}   \\
  EMEAC  &                       & \multicolumn{3}{l}{Equivalent machine EAC}              \\
  IMEAC  &                       & \multicolumn{3}{l}{Individual machine EAC}             
  \end{tabular}

\section{Introduction}

%
%
%
%
\subsection{LITERATURE REVIEW}
\raggedbottom
\IEEEPARstart{P}{ower} system is originally formed by multiple physical machines with complicated interactions. 
In order to analyze the transient stability of the entire power system, the system engineer focuses on the stability of the each machine in the system. 
In the transient stability analysis, the machine is modeled through its equation motion with its unique trajectory. The machine can be physically real or equivalent. 
Against the background that the power system is originally formed by multiple physically real machines, the individual machine analysts essentially focus on the transient behavior of each physically real individual machine in the system. 
The individual-machine thinking was originally formed by Fouad and Stanton \cite{1}, \cite{2}.
In the two paper series, they stated that ``the instability of a multi-machine system is determined by the motion of some unstable critical machines if more than one machine tends to lose synchronism”. Inspired by this idea, Vittal and Fouad stated that the instability of the system was decided by the individual-machine \cite{3}, \cite{4}. 
The individual-machine idea was further developed by Stanton in Refs. \cite{5}-\cite{7}. The discussions, conjectures and also hypotheses in these papers greatly inspire the future individual machine studies. Recently, systematical progress was shown in the individual-machine studies. 
The novel hybrid individual-machine-EAC method (IMEAC) was proposed \cite{8}-\cite{10}. This distinctive method was also supported by the individual-machine-transient-energy and individual-machine potential energy surface in a theoretical level \cite{11}, \cite{12}. Then, the mapping between the generalized Newtonian system and multi-machine power system is established \cite{13}. All these works originate the IMEAC method.
\par
 Compared with the individual machine angle that monitors the stability of each machine, another ``global” angle was also found in the power system transient stability. Global analysts focus on the ``energy superimposition” or ``motion equivalence” of all machines in the system. 
In particular, based on the global monitoring of the system trajectory, the IMTEs of all machines in the system are superimposed into the superimposed-machine transient energy (SMTE).
This energy superimposition originates the UEP method \cite{14} and the sustained fault method \cite{15}, \cite{16}. However, the legacy defect is that the NEC fails in SMTE because residual kinetic energy always exists at the critical energy point \cite{17}, \cite{18}.
In order to solve this problem, the equivalent machine was proposed. The equivalent machine can be seen as the milestone in the history of global methods. In this method, the motions of all machines are aggregated as two equivalent machines through the separations of two groups \cite{19}.
In this way the stability of the system is finally characterized precisely by using the equivalent-machine EAC (EMEAC), and the legacy residual SMKE problem is addressed. The EMEAC originates the IEEAC method \cite{19}.
\par
Revisiting the history of power system transient stability, individual machine, superimposed machine and equivalent machine can be seen as the three major perspectives in modern TSA. The thinking of each perspective is also clear and distinctive.
However, if we take a deep insight into the mechanisms of these transient stability methods, one can find that ``trajectory separation”, ``energy conversion” and ``EAC” were frequently found when using these methods in TSA. 
This fully reveals that the modern power transient stability analysis shows strong similarities with Newtonian system stability. In particular, the mechanism of the trajectory stability of a multi-machine system can be expressed as the altitude of the energy ball that rolls inside the basin \cite{13}.
The conversion between ``transient kinetic energy” and ``transient potential energy” is quite similar to the Newtonian energy conversion. The EAC can also be seen as the reflection of the Newtonian work \cite{13}. 
\par
The Newtonian-mechanics-based machine paradigms can be seen as the extractions of the ``common thinking” among the individual machine, superimposed machine and the equivalent machine. Compared with certain specific machine, the machine paradigms stand at a higher level.
This is because the advantages or even defects of the machines in TSA can be theoretically explained through these paradigms. In addition, historical conjectures, hypotheses and legacy problems will also be clarified through these paradigms.
\par
At the beginning of this six paper series, the authors emphasize that the transient stability paradigms cannot be seen as the ``fundamental theory” in the power system transient stability, because the fundamental theory of the power system transient stability is based on the trajectory stability. 
However, these paradigms can be used as a ``guidance” to take a deep insight into the mechanisms of the power system transient stability, and they can even be used as a ``tool” to develop novel transient stability methods.
\subsection{SCOPE AND CONTRIBUTION OF THE PAPER}

This six paper series can be seen as the explanations and clarifications of the fundamental mechanisms of individual machine, superimposed machine and the equivalent machine in TSA through machine paradigms.
This paper series require the authors already have a general understanding about the individual-machine \cite{8}-\cite{13}. Then, the superimposed machine \cite{21}, equivalent machine \cite{22}, inner-group machine \cite{23} and machine transformations \cite{24} will be explained in a genuine individual-machine manner \cite{20} in the companion papers.
\par In the first paper, the machine paradigms are developed according to the common thinking among the three methods. The machine paradigms comprise of trajectory paradigm, modeling paradigm and energy paradigm. The trajectory paradigm can be seen as the reflection of the trajectory stability; the modeling paradigm is the two-machine-system modeling of the trajectory stability; and the energy paradigm is the stability evaluation of the two-machine system.
The strict followings of these paradigms bring two advantages in TSA: (i) the trajectory stability of the machine is evaluated effectively at MPP, and (ii) the trajectory variance of the machine is depicted clearly at MPP. 
After that, the machine paradigms are naturally expressed into two forms, i.e., the individual-machine form and the equivalent machine transient form. Then, the relationship between the machine stability and the system stability are also analyzed. It is clarified that the stability of the original system is characterized in a ``machine-by-machine” manner, while the stability of the equivalent system is characterized in a ``one-and-only” equivalent machine manner. The machine paradigms essentially explains the effectiveness of both the individual-machine method and the equivalent-machine in TSA.
\par The contributions of this paper are summarized as follows:
\\
(i) Machine paradigms are established. This extracts the common thinking of the transient stability methods.
\\
(ii) The advantages of the machine paradigms are analyzed. This validates the use of the Newtonian mechanics in the power system transient stability.
\\
(iii) The individual-machine expression and the equivalent-machine expression of the machine paradigms are provided. This clarifies the effectiveness of both the individual-machine and the equivalent-machine in TSA.
\par The reminder of the paper is organized as follows. In Section \ref{section_II}, the Newtonian system is revisited. In Section \ref{section_III}, the Newtonian mechanics based machine paradigms are developed. In Section \ref{section_IV}, different roles of machine paradigms are analyzed. In Section \ref{section_V}, individual-machine expression is given. In Section \ref{section_VI}, the equivalent-machine expression is analyzed. In Section \ref{section_VII}, the effectiveness of both the individual-machine and the equivalent-machine are demonstrated. Conclusions are given in Section \ref{section_VIII}.

\section{BACKGROUND OF THE NEWTONIAN SYSTEM}\label{section_II}
\subsection{RELATIVE TRAJECTORY IN THE NEWTONIAN SYSTEM}

In this section, the Newtonian energy conversion is revisited first \cite{13}. This revisit may enhance the readers to understand the mapping between the power system transient stability and Newtonian stability. Note that this revisit is reorganized in the ``trajectory of the ball”, ``two-ball system modeling” and ``Newtonian energy conversion” order.
\\
\textit{Equation of motion}: In the classic Newtonian mechanics, the Newtonian system is formed by the two balls, i.e., the ball and the Earth. Each ball has its own equation of motion. The motions of the two balls are depicted as
\begin{equation}
  \label{equ1}
  \normalsize
    \left\{\begin{array}{l}
      \frac{d h_{\mathrm {ball }}}{d t}=v_{\mathrm {ball }} \\
      \\
      m_{\mathrm {ball }} \frac{d v_{\mathrm {ball }}}{d t}=F_{\mathrm {ball }}
      \end{array}\right.
      \left\{\begin{array}{l}
        \frac{d h_{\mathrm {Earth }}}{d t}=v_{\mathrm {Earth }} \\
        \\
        m_{\mathrm {Earth }} \frac{d v_{\mathrm {Eath }}}{d t}=F_{\mathrm {Earth }}
        \end{array}\right.
\end{equation}
\par
\begin{figure}[H]
  \centering
  \includegraphics[width=0.3\textwidth,center]{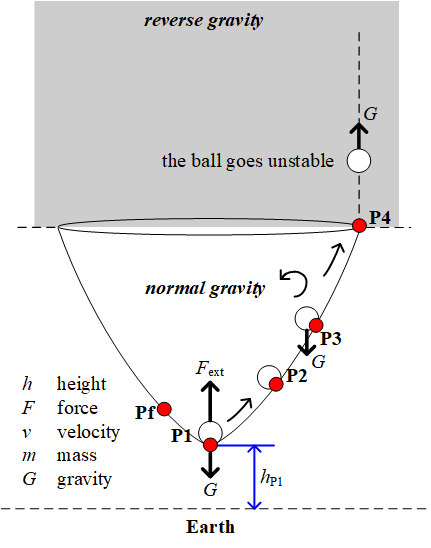}
  \caption{Structure of the original Newtonian system \cite{13}.} 
  \label{fig1}   
\end{figure}

 In Eq. (\ref{equ1}), $F_\text{net}$ and \textit{h} are the ``net force" and the altitude of the ball, respectively.
$F_\text{Earth}$ and $h_\text{Earth}$ are the ``net force” and the altitude of the Earth, respectively. Other parameters are shown in Fig. \ref{fig1}.
\par Because the Earth maintains stationary ($m_\text{earth}=\infty$, $h_\text{Earth}=0$), The equation of motion of the Earth is given as
\begin{equation}
  \label{equ2}
  \left\{\begin{array}{l}
    \frac{d h_{\mathrm {Earth }}}{d t}=v_{\mathrm {Earth }}=0 \\
    \\
    \frac{d v_{\mathrm {Eath }}}{d t}=\frac{F_{\mathrm {Earth }}}{m_{\mathrm {Eath }}}=0
    \end{array}\right.
\end{equation}
\par Following Eq. (\ref{equ2}), the stationary Earth is set as the ``motion reference'' of the system.
\par The structure of the original Newtonian system is shown in Fig. \ref{fig1}. Note that the original Newtonian system is defined as a “gravity-variant” system \cite{13}. That is, the gravity of the ball will become reversed once the ball goes through the boundary of the basin.

\par In the Newtonian system, at first, the ball stays at the bottom of the basin (P1), as shown in Fig. \ref{fig4}.
Then, at $t_\text{0}$, an ``external force'', i.e., the disturbance imposes on the ball. Once the ball reaches P2, the external force is cleared at the moment. 
After that, the only force that imposes on the ball becomes the ``internal force'', i.e., the gravity. The ball starts decelerating with only gravity on it, 
and it rolls towards the boundary of the basin.
\\
\textit{Trajectory of the ball}: Using the Earth as the motion reference, the trajectory of the ball is defined as
\begin{equation}
  \label{equ3}
  h=h_{\mathrm {ball }}-h_{\mathrm {Earth }}=h_{\mathrm {ball }}
\end{equation}
\par
\begin{figure}[H]
  \centering
  \includegraphics[width=0.45\textwidth,center]{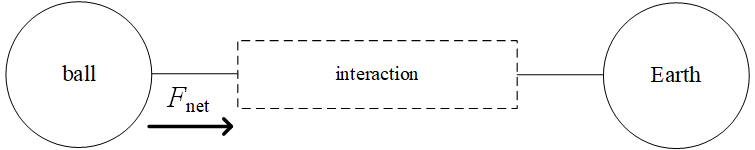}
  \caption{Formation of the original Newtonian system..} 
  \label{fig2}   
\end{figure}

 In Eq. (\ref{equ3}), \textit{h} depicts the separation of the ball with respect to the Earth.
\\
\textit{Two-ball system modeling}: Based on the two equation of motions as given in Eqs. (\ref{equ1}) and (\ref{equ2}), a two-ball system, i.e., the Newtonian system is formed by the ball and the Earth. The formation of the Newtonian system is shown in Fig. \ref{fig2}.
\par The modeling of the Newtonian system is given as
\begin{equation}
  \label{equ4}
  \left\{\begin{array}{l}
    \frac{d\left(h_{\mathrm {ball }}-h_{\mathrm {Earth }}\right)}{d t}=v_{\mathrm {ball }}-v_{\mathrm {Earth }}=v_{\mathrm {ball }} \\
    \\
    m_{\mathrm {ball }} \frac{d\left(v_{\mathrm {ball }}-v_{\mathrm {Eath }}\right)}{d t}=F_{\mathrm {ball }}-\frac{m_{\mathrm {ball }}}{m_{\mathrm {Eath }}} F_{\mathrm {Eath }}=F_{\mathrm {ball }}
    \end{array}\right.
\end{equation}
\par From Eq. (\ref{equ4}), the trajectory of the ball by using the Earth as the motion reference ($h_\text{ball}$) is modeled through the two-ball based Newtonian system.

\subsection{NEWTONIAN ENERGY CONVERSION}
\textit{Newtonian energy definition}: Assume the energy reference point of the system is set as P1. Based on the two-ball Newtonian system as Eq. (\ref{equ4}), the Newtonian total energy (TE) of the ball is denoted as
\begin{equation}
  \label{equ5}
  V^{(\mathrm{P})}=\frac{1}{2} m v^{(\mathrm{P}) 2}+\int_{h^{(\mathrm{P} 1)}}^{h^{(\mathrm{P})}} G d h
\end{equation}
\par In Eq. (\ref{equ5}), TE comprises of two components, i.e., the KE and the PE.
\par Following the definition of the Newtonian energy, the distinctive ``maximum potential energy point” (MPP) can be used for stability characterization.
In the Newtonian system, both DSP and DLP are the MPPs. Detailed analysis is given as below
\\
\textit{Dynamic liberation point (DLP)}: Once the ball accumulates very high KE at P2 after disturbance, this KE cannot be absorbed inside the basin.
Then, the ball will go through the boundary at P4, as in Fig. \ref{fig4}. P4 is named the  ``dynamic liberation point” (DLP). After that, the ball falls into the reverse gravitational field and accelerates again.
Against this background, DLP depicts the point where the ball becomes unstable. The occurrence of DLP is depicted as
\begin{equation}
  \label{equ6}
  G_\text{DLP}=0
\end{equation}
\textit{Dynamic stationary point (DSP)}: Compared with unstable case, if the ball accumulates low KE at P2 after disturbance, this KE will be completely absorbed before reaching DLP. 
Then, the ball inflects back at P3, as in Fig. \ref{fig1}. P3 is named the ``dynamic stationary point” (DSP).
Against this background, DSP depicts the point where the ball maintains stable. The occurrence of DSP is depicted as
\begin{equation}
  \label{equ7}
  v_\text{DSP}=0
\end{equation}
\textit{Characteristics of the Newtonian energy conversion (NEC)}: Following the characteristics of DSP and DLP, the distinctive characteristic of the Newtonian system is that the stability of the system can be characterized effectively through the residual KE at MPP \cite{13}.
\par The residual KE at MPP is denoted as
\begin{equation}
  \label{equ8}
  V_{K E}^{R E}=V_{K E}^{(\mathrm{P} 2)}-\Delta V_{P E}
\end{equation}
where
\begin{spacing}{1.5}
 \noindent$V_{K E}^{(\mathrm{P} 2)}=\frac{1}{2} m v^{(\mathrm{P} 2) 2}$
  \\$\Delta V_{P E}=\int_{h^{(P 1)}}^{h^{\mathrm{MPP}}} G d h-\int_{h^{(\mathrm{Pl})}}^{h^{(\mathrm{P 2})}} G d h=\int_{h^{(\mathrm{P} 2)}}^{h^{\mathrm{MPP}}} G d h$ 
\end{spacing}
In Eq.(\ref{equ8}), $V_{K E}^{R E}$ is the residual KE at the MPP. The ball will become unstable if $V_{K E}^{R E}$ is positive at MPP (the MPP would be the DLP);
and the ball will maintain stable if $V_{K E}^{R E}$ is zero at MPP (the MPP would be the DSP).
\par From analysis above, clear energy conversion between KE and PE can be found in the original Newtonian system. This energy conversion is also named the ``Newtonian energy conversion” (NEC).
The ball would go unstable if the residual KE occurs at MPP \cite{13}. 
\\
\textit{Equal Area Criterion (EAC)}: In the \textit{h}-\textit{F} space, the EAC can be seen as the Newtonian work, and it is proved to be identical to NEC \cite{13}.
Based on this, the residual KE can also be depicted as
\begin{equation}
  \label{equ9}
  V_{K E}^{R E}=A_\mathrm{acc}-A_\mathrm{dec}
\end{equation}
where
\begin{spacing}{1.5}
  \noindent$A_\mathrm{acc}=V_{KE}^{(\mathrm{P} 2)}$\\
   $A_{\mathrm{dec}}=\int_{h^{(\mathrm{P} 2)}}^{h^{\mathrm{MPP}}} G d h$
\end{spacing}
From Eq. (\ref{equ9}), the ball would become unstable if the acceleration area is larger than the deceleration area \cite{13}. Therefore, NEC and EAC are completely identical \cite{13}.
\subsection{STABILITY CHARACTERIZATION OF THE NEWTONIAN SYSTEM} \label{section_IIC}
\noindent\textit{Stability characterization}: The stability characterizations of the original Newtonian system are given as below.
\\
\\
(i) From the perspective of NEC, the system is evaluated to go unstable if the residual KE occurs at MPP.
\\
(ii) From the perspective of EAC, the system is evaluated to go unstable if the acceleration area is larger than the deceleration area.
\\
\par Note that (i) and (ii) are completely identical.
\\
\textit{Trajectory depiction}: The trajectory depictions of the ball using MPP are given as below
\\
\\
(i) DLP is the point where the trajectory of the ball starts separating ($\mathrm{d}^{2} h / \mathrm{d} t^{2}=G_{\mathrm{DLP}}=0$).
\\
(ii) DSP is the point where the trajectory of the ball inflects back ($\mathrm{d}^ h / \mathrm{d} t=v_{\mathrm{DLP}}=0$).
\\
\par (i) and (ii) indicate that the trajectory variance can be depicted clearly through DLP or DSP. In fact, DLP and DSP is just the reflection of the NEC in the trajectory.
\subsection{ADVANTAGES OF NEC}
Following the analysis in Section \ref{section_IIC}, the two advantages of the NEC in the stability analysis of the Newtonian system are summarized as below
\\
\par\textit{Stability-characterization advantage: The trajectory stability of the ball is characterized precisely at MPP}.
\par\textit{Trajectory-depiction advantage: The trajectory variance of the ball is depicted clearly at MPP}.
\\
\par
\begin{figure}[H]
  \centering
  \includegraphics[width=0.42\textwidth,center]{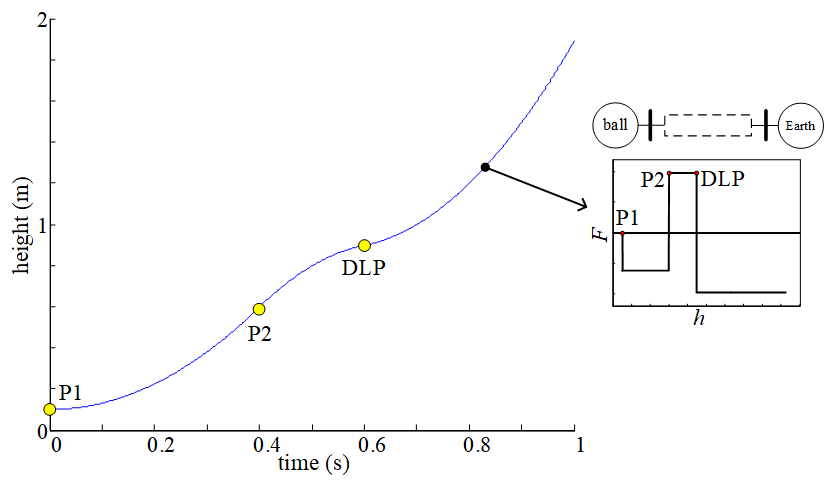}
  \caption{Trajectory of the ball \cite{13}.} 
  \label{fig3}   
\end{figure}

\begin{figure}[H]
  \centering
  \includegraphics[width=0.34\textwidth,center]{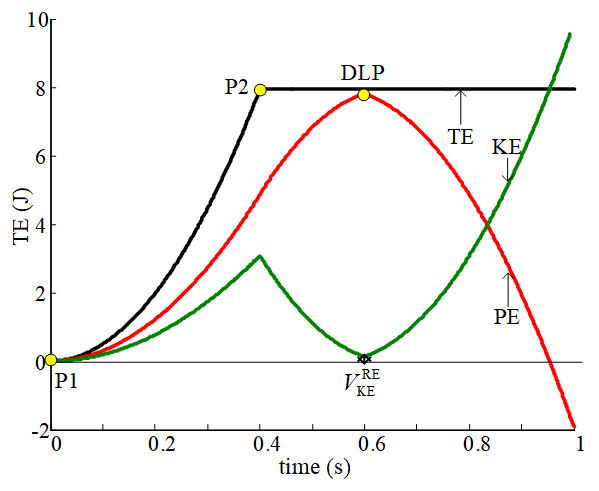}
  \caption{NEC in the unstable original Newtonian system \cite{13}.} 
  \label{fig4}   
\end{figure}

 Demonstrations about the two advantages are given below. Following the case in Ref. \cite{13}, the trajectory of the ball along time horizon by using the Earth as the motion reference is shown in Fig. \ref{fig3}
The The ball-Earth system and the EAC is also shown the figure. The NEC of the Newtonian system is shown in Fig. \ref{fig4}. 
From Figs. \ref{fig3} and \ref{fig4}, the trajectory stability of the ball is evaluated effectively at DLP, and the trajectory separation of the ball is depicted clearly through DLP.
\subsection{EQUIVALENCE BETWEEN TRAJECTORY STABILITY AND NEWTONIAN STABILITY}
If we take a deep insight into the trajectory stability and NEC of the ball, one can obtain the following
\\
\par\textit{The trajectory stability is identical to the Newtonian stability}.
\\
\par In particular, the trajectory of the ball will go infinite along time horizon only when the ball goes through DLP and it falls in the reversed gravitational field; The trajectory of the ball will become bounded only when the ball inflects back at DSP and it rolls inside the basin.
\par
\begin{figure}[H]
  \centering
  \includegraphics[width=0.42\textwidth,center]{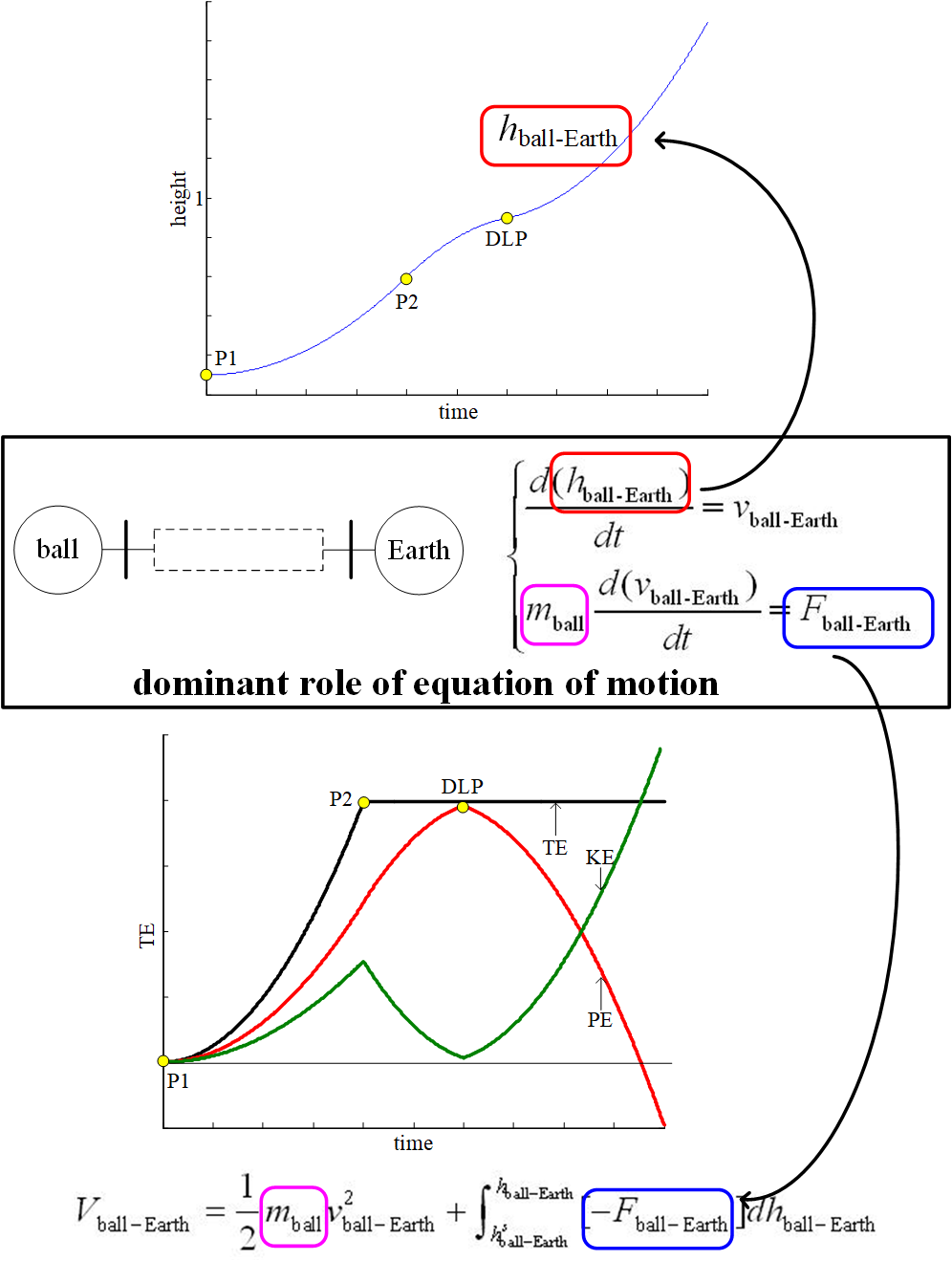}
  \caption{Equivalence between trajectory stability and Newtonian stability.} 
  \label{fig5}   
\end{figure}

 The equivalence between the trajectory stability and Newtonian energy stability is shown in Fig. \ref{fig5}.
From Fig. \ref{fig5}, the trajectory is modeled through the two-ball based Newtonian system, and the energy is defined based on the two-ball system. Therefore, the equivalence between trajectory and NEC is fully ensured by the ``bridge” between them, i.e., the equation of motion of the Newtonian system. Based on this ``bridge”, the following holds
\\
\par\textit{Strict correlation between trajectory and energy is established based on the two-ball modeling of the Newtonian system}.
\\
\par This also explains the reason why NEC brings the two distinctive advantages in the stability characterization and trajectory depiction.
\par We go a further step. if a Newtonian-like physical system shows strict mappings with the original Newtonian system, then one can obtain the following
\\
\par\textit{The advantages of NEC will be inherited in this Newtonian-like system}.
\\
\par In particular, the stability of this Newtonian-like system can be characterized precisely at MPP, and the trajectory variance of this Newtonian-like system can be depicted clearly at MPP.
\par Following the analysis in Ref. \cite{13}, the multi-machine power system is a typical Newtonian-like system because strict mappings are found between the two systems. Against this background, the advantages of NEC will be fully inherited in the TSA if it follows the Newtonian mechanics based stability analysis, i.e., the trajectory monitoring, two-ball system modeling and NEC (EAC) in sequence.

\section{NEWTONIAN MECHANICS BASED TRANSIENT STABILITY PARADIGMS} \label{section_III}
\subsection{TRAJECTORY PARADIGM} \label{section_IIIA}
Following the analysis in Section \ref{section_II}, the Newtonian-mechanics, machine paradigms comprise of the three elements, i.e., trajectory paradigm, modeling paradigm and the energy paradigm. Note that the analysis in this section strictly follows the structure as given in Section \ref{section_II}.
\par In the synchronous reference, the equation of motion of each machine is denoted as
\begin{equation}
  \label{equ10}
  \left\{\begin{array}{l}
    \frac{d \delta_{i}}{d t}=\omega_{i} \\
    \\
    M_{i} \frac{d \omega_{i}}{d t}=P_{i}
    \end{array}\right.
\end{equation}
\par In Eq. (\ref{equ10}), $\delta_{i}$ and $P_{i}$ are the rotor angle and the net force of the machine, respectively. Other parameters are already given in Ref. \cite{8}.
$\delta_{i}$ is named the ``trajectory” of the machine along time horizon. Note that the machine can be individual or equivalent.
\par For a multi-machine power system that suffers severe disturbance, the balance inside each machine will be destroyed. After that, the trajectories of some machines might separate along time horizon. Under this background, the trajectory instability will occur and the system becomes unstable. Therefore, the power system transient stability should be explicitly depicted as the power system ``trajectory stability”.
\par In order to depict the trajectory separations among machines in the system, generally, one machine is chosen as the ``motion reference” of all the other machines in the system. This machine is also named the``reference machine” (RM). Based on this, the motions of all the other machines in the system become ``relative” with respective to the RM.
\par In the RM reference, the relative trajectory of each objective machine is denoted as
\begin{equation}
  \label{equ11}
  \delta_{i\mbox{-}\mathrm{RM}}=\delta_{i}-\delta_{\mathrm{RM}}
\end{equation} 
\par In Eq. (\ref{equ11}), RM can be individual or equivalent \cite{8}. Machine \textit{i} is named the ``objective machine” in the RM reference. From the equation, $\delta_{i\mbox{-}\mathrm{RM}}$ represents the separation of the objective machine with respect to the RM. Note that the trajectory of the RM will become a zero horizontal line in the RM reference.
\par Based on the analysis above, the first machine paradigm, i.e., the trajectory paradigm is given as
\\
\par \textit{Trajectory paradigm: an objective machine becoming unstable is depicted as the trajectory of the machine going infinite in the RM reference}.
\\
\par
\begin{figure}[H]
  \centering
    \subfigure[]{  
      \includegraphics[width=0.38\textwidth,center]{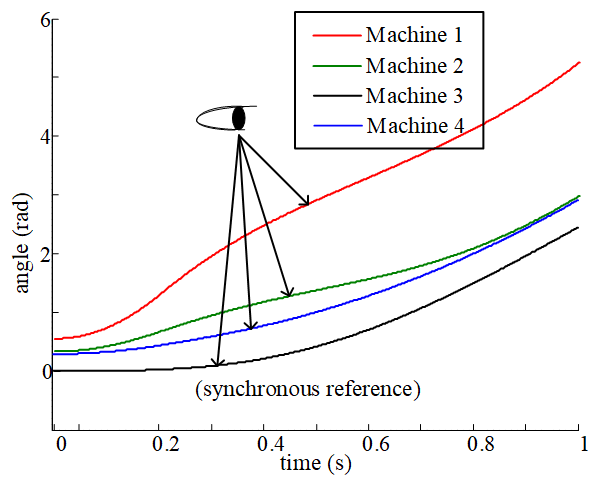}}    
      \centering
      \subfigure[]{
        \includegraphics[width=0.38\textwidth,center]{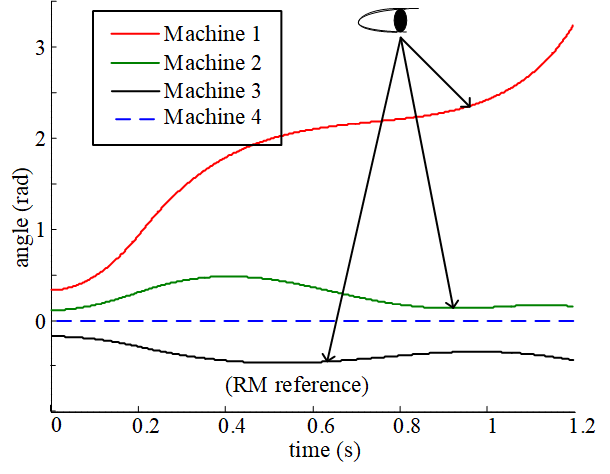}}     
   \caption{Trajectory description of each machine. (a) Synchronous reference. (b) RM reference.}
  \label{fig6}
\end{figure}

 A tutorial example is given below to demonstrate the trajectory paradigm. Assume four machines are given, and these machines can be individual or equivalent. 
The trajectory monitoring in the synchronous reference and that in the RM reference are shown in Figs. \ref{fig6} (a) and (b), respectively.
\par From Fig. \ref{fig6}, three trajectories of objectives machines are formed in the RM reference. The system engineer monitors the trajectory of each machine in parallel. Along time horizon, the trajectory of Machine 1 goes infinite and it becomes unstable. Comparatively, the trajectories of Machines 2 and 3 oscillate and they both maintain stable. Following trajectory stability theory, since the trajectory separation occurs between Machine 1 and RM, Machine 1 becomes unstable and thus the system also becomes unstable.
\par Theoretically, an objective machine can be evaluated to go unstable if its trajectory in the RM reference goes ``infinite” along time horizon. However, technically, this will show very low efficiency in TSA. The two problems are given below
\\\par (i) The trajectory stability of the machine cannot be characterized effectively.
\par (ii) The trajectory variance of the machine cannot be depicted clearly.
\\\par In order to make the stability characterization and trajectory depiction of the machine become flexible, the ``modeling” of the trajectory variance should be established first. This originates the second machine paradigm, i.e., the modeling paradigm in TSA.
\subsection{MODELING PARADIGM} \label{section_IIIB}
The equations of motions of the objective machine and that of the RM in the synchronous reference are denoted as
\begin{equation}
  \left\{\begin{array} { l } 
    { \frac { d \delta _ { i } } { d t } = \omega _ { i } } \\
    \\
    { M _ { i } \frac { d \omega _ { i } } { d t } = P _ { i } }
    \end{array} \quad \left\{\begin{array}{l}
    \frac{d \delta_{\mathrm{RM}}}{d t}=\omega_{\mathrm{RM}} \\
    \\
    M_{\mathrm{RM}} \frac{d \omega_{\mathrm{RM}}}{d t}=P_{\mathrm{RM}}
    \end{array}\right.\right.
    \label{equ12}
\end{equation}
\par
\begin{figure}[H]
  \centering
  \includegraphics[width=0.45\textwidth,center]{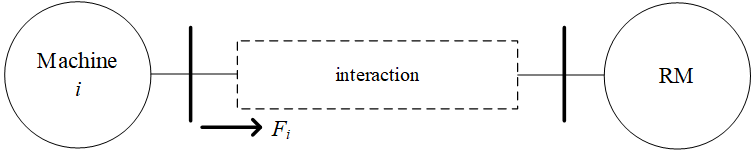}
  \caption{Two-machine system.} 
  \label{fig7}   
\end{figure}

 Based on Eq. (\ref{equ12}), a two-machine system is formed by the two machines. The formation of the objective-machine-RM system (O-RM) system is shown in Fig. \ref{fig7}.
Based on this formation, the modeling of the two-machine system is given as
\begin{equation}
  \left\{\begin{array}{l}
    \frac{d \delta_{i\mbox{-}\mathrm{RM}}}{d t}=\omega_{i\mbox{-}\mathrm{RM}} \\
    \\
    M_{i} \frac{d \omega_{i\mbox{-}\mathrm{RM}}}{d t}=f_{i\mbox{-}\mathrm{RM}}
    \end{array}\right. 
    \label{equ13}
\end{equation}
where
\begin{spacing}{1.5}
  \noindent$\delta_{i\mbox{-}\mathrm{RM}}=\delta_{i}-\delta_{\mathrm{RM}}$\\
  $\omega_{i\mbox{-}\mathrm{RM}}=\omega_{i}-\omega_{\mathrm{RM}}$\\
   $f_{i\mbox{-}\mathrm{RM}}=P_{i}-\frac{M_{i}}{M_{\mathrm{RM}}} P_{\mathrm{RM}}$
\end{spacing}
In Eq. (\ref{equ13}), $f_{i\mbox{-}\mathrm{RM}}$ reflects the interactions between the machine and the RM. From the equation, the trajectory variance of the machine in the RM reference ($\delta_{i\mbox{-}\mathrm{RM}}$) as in Fig. \ref{fig6} (b) is completely modeled through the two-machine system. 
\par Based on the analysis above, the second machine paradigm, i.e., the modeling paradigm is given as
\\
\par \textit{Modeling paradigm: The trajectory stability of the machine is modeled through the two-machine system}.
\\
\par
\begin{figure}[H]
  \centering
  \includegraphics[width=0.38\textwidth,center]{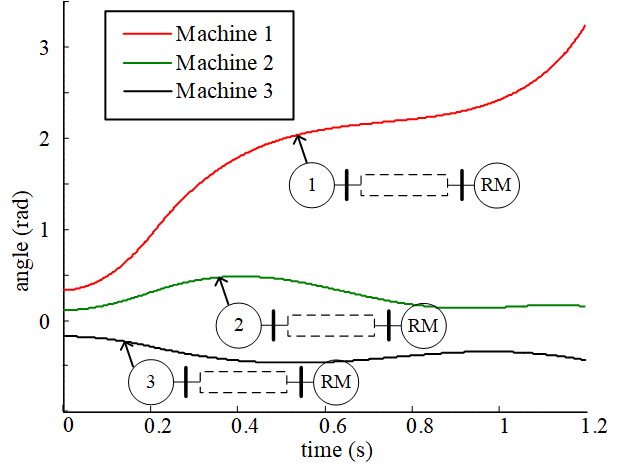}
  \caption{Formations of the two-machine systems.} 
  \label{fig8}   
\end{figure}

 For the tutorial example as given in Fig. \ref{fig6}, the formations of the two-machine systems are shown in Fig. \ref{fig8}.
From Fig. \ref{fig8}, the trajectory variance of each machine in the RM reference is modeled through its corresponding two-machine system. Against this background, three two-machine systems are formed. Note that RM is the same motion reference for all the three two-machine systems.
\par We go a further step. Based on the strict mapping between the Newtonian system and the two-machine-system as analyzed in Ref. \cite{13}, the ``Newtonian energy conversion” (NEC) can be used in the stability characterization of the two-machine system. This originates the third machine paradigm, i.e., the energy paradigm in TSA.

\subsection{ENERGY PARADIGM} \label{section_IIIC}
Following the definition of the Newtonian energy as given in Section \ref{section_II}, the  transient energy of Machine \textit{i} is defined as
\begin{equation}
  V_{i\mbox{-}\mathrm{RM}}=V_{K E i\mbox{-}\mathrm{RM}}+V_{P E i\mbox{-}\mathrm{RM}}
  \label{equ14}
\end{equation}
where
\begin{spacing}{1.5}
  \noindent$V_{KEi-\mathrm{RM}}=\frac{1}{2} M_{i} \omega_{i-\mathrm{RM}}^{2}$\\
  $V_{P E i-\mathrm{RM}}=\int_{\delta_{i-\mathrm{RM}}^{s}}^{\delta_{i-\mathrm{RM}}}\left[-f_{i-\mathrm{RM}}^{(P F)}\right] d \delta_{i-\mathrm{RM}}$
\end{spacing}
The residual KE of Machine \textit{i} at its MPP is denoted as
\begin{equation}
  \begin{split}
    \label{equ15}
    V_{K E i\mbox{-}\mathrm{RM}}^{R E}&=V_{K E i\mbox{-}\mathrm{RM}}^{c}-\Delta V_{P E i\mbox{-}\mathrm{RM}}\\
    &=A_{A C C i\mbox{-}\mathrm{RM}}-A_{DECi\mbox{-}\mathrm{RM}} 
  \end{split}
\end{equation}
where
\begin{spacing}{2}
  \noindent$V_{K E i\mbox{-}\mathrm{RM}}^{c}=\frac{1}{2} M_{i} \omega_{i\mbox{-}\mathrm{RM}}^{c 2}=A_{A C C i\mbox{-}\mathrm{RM}}$ \\
  \noindent$\Delta V_{P E i\mbox{-}\mathrm{RM}}=\int_{\delta_{i\mbox{-}\mathrm{RM}}^{S}}^{\delta_{i-\mathrm{RM}}^{MPP}}\left[-f_{i\mbox{-}\mathrm{RM}}^{(P F)}\right] d \delta_{i\mbox{-}\mathrm{RM}}-\\
  \int_{\bar{\theta}_{i\mbox{-}\mathrm{RM}}^{s}}^{\delta_{i\mbox{-}\mathrm{RM}}^{c}}\left[-f_{i\mbox{-}\mathrm{RM}}^{(P F)}\right] d \delta_{i\mbox{-}\mathrm{RM}}
  =A_{DECi\mbox{-}\mathrm{RM}}$
\end{spacing}
Following this Newtonian energy definition, the stability characterizations of the objective machine in the RM reference are summarized as below.
\\
\\
(i) From the perspective of transient energy conversion, the machine is evaluated to go unstable if the residual KE occurs at its MPP.
\\
(ii) From the perspective of EAC, the machine is evaluated to go unstable if the acceleration area is larger than the deceleration area.
\par (i) and (ii) are completely identical.
\par Following all the analysis above, the energy paradigm is given as
\\
\par \textit{Energy paradigm: The NEC (EAC) is used as the stability characterization of the two-machine system}.
\\\par The two advantages of individual-machine transient energy conversion is given as
\\
\\Stability-characterization-advantage: The stability of the machine is characterized precisely at MPP.
\\Trajectory-advantage: The trajectory of the machine is depicted clearly at MPP.
\\\par
\begin{figure}[H]
  \centering
  \includegraphics[width=0.4\textwidth,center]{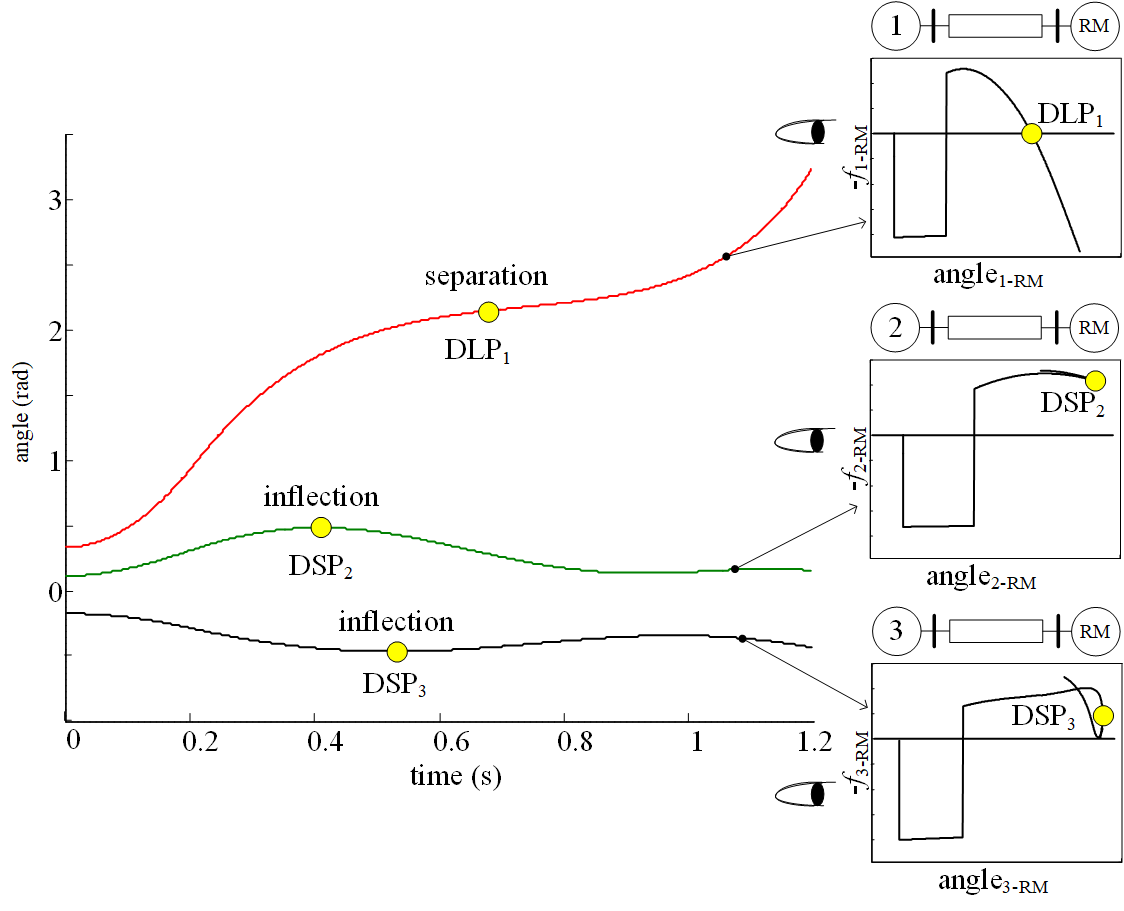}
  \caption{Trajectory and corresponding EAC inside each objective machine.} 
  \label{fig9}   
\end{figure}

 A tutorial example about the energy paradigm is demonstrated in Fig. \ref{fig9}.
From Fig. \ref{fig9}, the stability of each two-machine system is characterized precisely by using EAC. Based on the strict mappings between the Newtonian system and the multi-machine power system \cite{13}, the two advantages of NEC will be completely inherited in transient stability analysis.
In particular, the trajectory stability of each objective machine is modeled through the two-machine system, and the stability of the two-machine system is evaluated effectively through NEC. Against this background, the trajectory stability of each objective machine is characterized precisely at MPP, and the trajectory variance of each objective machine is depicted clearly at MPP.

\subsection{FURTHER ANALYSIS} \label{section_IIID}
Following the analysis of all the paradigms above, tutorial explanations of the machine paradigms are given as below
\par (i) Because the transient stability is originally a trajectory stability problem, this trajectory stability is depicted through the trajectory separation between the objective machine and the RM.
\par (ii) Because the trajectory separation shows low efficiency in the stability evaluation, the trajectory variance is modeled through the two-machine system.
\par (iii) Because strict mapping is established between the Newtonian system and the two-machine system, the stability of the two-machine system is finally characterized through transient energy that is defined in a Newtonian form.
\par Following (i) to (iii), the use of the machine paradigms in TSA can be expressed as
\\\par \textit{Machine paradigms establish the strict correlation between trajectory and energy of the objective machine through the two-machine system modeling}.
\\\par In particular, if the structure of an objective machine strictly follows the machine mechanisms, the strict correlation between trajectory and energy will be established. Further, the establishment of this strict correlation fully ensures the stability characterization advantage and the trajectory depiction advantage of the machine. In fact, machine paradigms can be seen as the use of the Newtonian mechanics in the power system transient stability analysis.
\par The deduction above can also be given in a contrary way
\\\par \textit{The machine will become pseudo without equation of motion if it violates the machine paradigms}.
\\\par In particular, if a machine violates the machine paradigms, the machine will become a pseudo machine because it does not have its corresponding equation of motion. That is, the pseudo machine does not have trajectory, mass and force on it. Against this background, the pseudo machine becomes ``meaningless” in TSA.

\section{DIFFERENT ROLES OF THE MACHINE PARADIGMS} \label{section_IV}

\subsection{THE PRIMARY ROLE OF THE TRAJECTORY PARADIGM} \label{section_IVA}
Among all the three paradigms, it is clear that the trajectory paradigm is the complete reflection of the original definition of the trajectory stability. Therefore, the trajectory paradigm is also the primary paradigm among all the three paradigms. Based on this, one can naturally obtain the following
\\\par \textit{The trajectory monitoring essentially decides the formation of the two-machine system and its corresponding transient energy conversion}.
\\\par Originated from the trajectory paradigm, two expressions of the system trajectory can be naturally formed. That is, the ``individual machine” expression and the ``equivalent machine” expression. Detailed analysis is given in Section \ref{section_V}.

\subsection{THE DOMINANT ROLE OF THE MODELING PARADIGM} \label{section_IVB}
Following the machine paradigms as analyzed in Section \ref{section_III}, ``trajectory” is simulated based on the two-machine system with the relative equation of motion, and ``transient energy” is defined based on the two-machine system with the relative equation of motion in a Newtonian energy manner.
\par Based on the analysis above, an extreme question can be emerged: which element is ``unnecessary” in the power system transient stability?
\par The answer should be the energy. This is because the power system transient stability is originally defined as a trajectory stability problem. In other words, the machine can be identified as unstable only through its trajectory becoming infinite along time horizon.
The energy is just a ``definition” that is used to measure the trajectory variance. Against this background, although the energy show advantages in both stability characterization and trajectory depiction, it is ``unnecessary” in the power system transient stability.
\par Following the fact that the energy is unnecessary, another companion question can also be emerged: which element is ``dominant” in the power system transient stability?
\par It is clear that this element should be the two-machine system modeling that is physically depicted as the relative equation of motion. The reason is given as below
\\\par \textit{The trajectory variance is modeled through the two-machine system}.
\par \textit{The transient energy conversion is also defined based on the two-machine system}.
\\\par In fact, the most crucial concept in the power system transient stability, i.e., the trajectory must be simulated based on the relative equation of motion of the two-machine system. The modeling paradigm is the ``bridge” between the trajectory paradigm and the energy paradigm.
\par
\begin{figure}[H]
  \centering
  \includegraphics[width=0.45\textwidth,center]{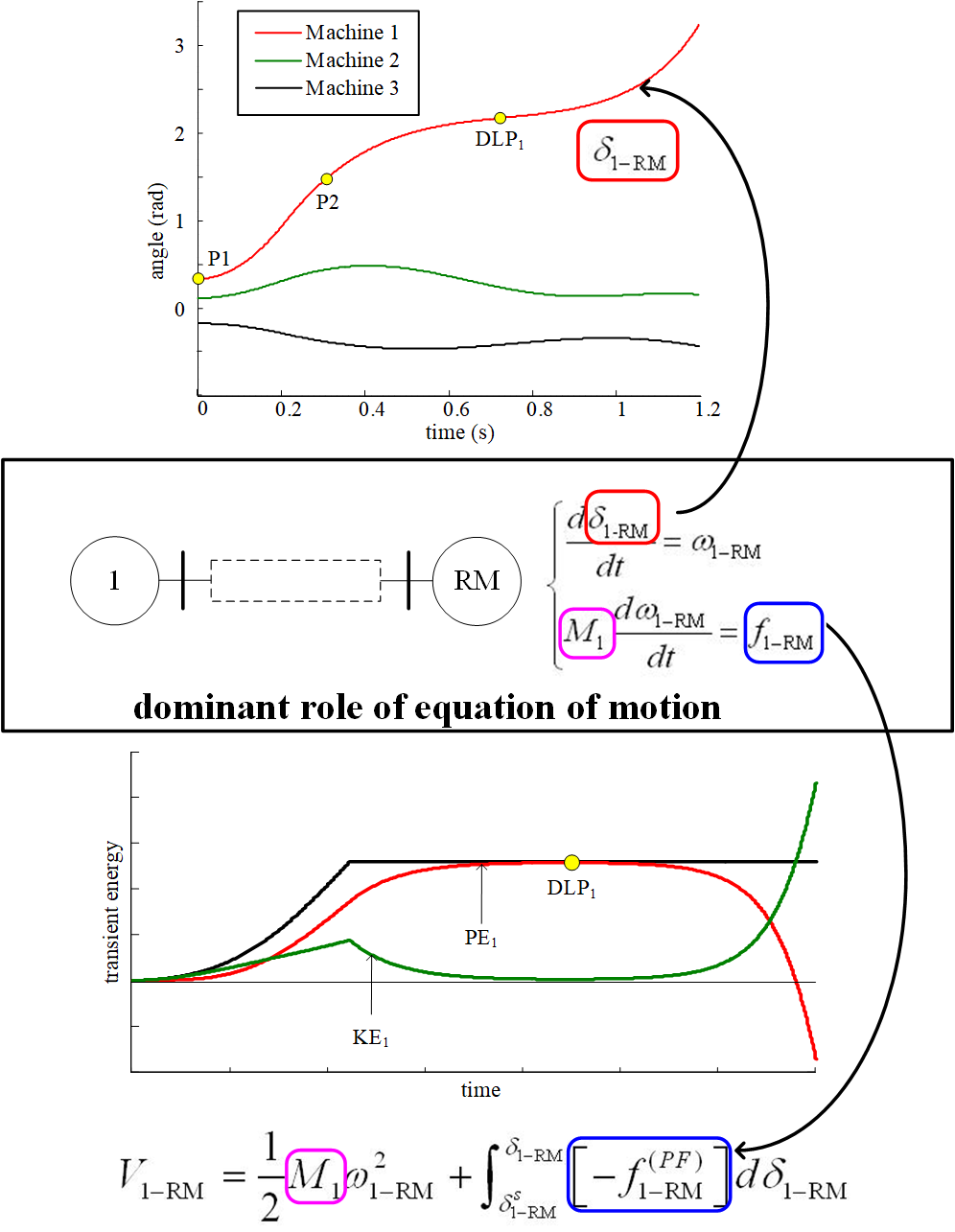}
  \caption{Correlation between trajectory and energy through two-machine system modeling.} 
  \label{fig10}   
\end{figure}

 The correlation between the trajectory stability and energy stability through two-machine system modeling is shown in Fig. \ref{fig10}.
From Fig. \ref{fig10}, it is the modeling paradigm that ensures the strict correlation between trajectory and energy. This strict correlation further ensures the two advantages in TSA as given in Section \ref{section_IIIC}.
Note that the two advantages hold simultaneously because they both are based on the modeling of the two-machine system.
\par In the following section, following the primary role of the trajectory paradigm, the individual-machine expression and equivalent machine expression are provided. This may help readers visually understand the mechanism of the individual machine and equivalent machine that both strictly follow the machine paradigms in TSA.
In addition, the following analysis is given in tutorial. That is, the transient energy of the two machines is given in an EAC manner. Detailed NEC characteristics of the two machines will be analyzed in the companion papers.

\section{INDIVIDUAL-MACHINE EXPRESSION} \label{section_V}

\subsection{INDIVIDUAL-MACHINE TRAJECTORY} \label{section_VA}
\noindent \textit{Equation of motion}: In the power system transient stability analysis, the ``real” machine is the only component that physically exists in a multi-machine system. That is, the original multi-machine system trajectory is formed by the trajectory of each ``real” machine in the synchronous reference.
\par The equation of motion of each physically real machine in the synchronous reference is denoted as
\begin{equation}
  \left\{\begin{array}{l}
    \frac{d \delta_{i}}{d t}=\omega_{i} \\
    \\
    M_{i} \frac{d \omega_{i}}{d t}=P_{m i}-P_{e i}
    \end{array}\right. 
    \label{equ16}
\end{equation}
\par In Eq. (\ref{equ16}), all the parameters are given in Ref. \cite{6}.
\par In order to measure the trajectory separation among all machines in the system, the equivalent Machine-SYS is developed, and it is used as the RM. The equation of motion of Machine-SYS is denoted as
\begin{equation}
  \left\{\begin{array}{l}
    \frac{d \delta_{\mathrm{SYS}}}{d t}=\omega_{\mathrm{SYS}} \\
    \\
    M_{\mathrm{SYS}} \frac{d \omega_{\mathrm{SYS}}}{d t}=P_{\mathrm{SYS}}
    \end{array}\right. 
    \label{equ17}
\end{equation}
where
\begin{spacing}{1.5}
  \noindent$M_{\mathrm{SYS}}=\sum_{i=1}^{n} M_{i}$\\
  $\delta_{\mathrm{SYS}}=\frac{1}{M_{\mathrm{SYS}}} \sum_{i=1}^{n} M_{i} \delta_{i}$\\
  $\omega_{\mathrm{SYS}}=\frac{1}{M_{\mathrm{SYS}}} \sum_{i=1}^{n} M_{i} \omega_{i}$\\
  $P_{\mathrm{SYS}}=\sum_{i=1}^{n}\left(P_{m i}-P_{e i}\right)$  
\end{spacing}

\par From Eq. (\ref{equ17}), the motion of Machine-SYS represents the equivalent motion of all machines in the system.\\
\textit{Trajectory depiction}: Following trajectory paradigm, in the COI-SYS reference, the trajectory of each machine is denoted as
\begin{equation}
  \delta_{i\mbox{-}\mathrm{SYS}}=\delta_{i}-\delta_{\mathrm{SYS}}
    \label{equ18}
\end{equation}
\par In Eq. (\ref{equ18}), $\delta_{i\mbox{-}\mathrm{SYS}}$ going infinite along time horizon would indicate that the machine becomes unstable.
$\delta_{i\mbox{-}\mathrm{SYS}}$ is named the individual-machine trajectory (IMTR).
\par
\begin{figure}[H]
  \centering
  \includegraphics[width=0.4\textwidth,center]{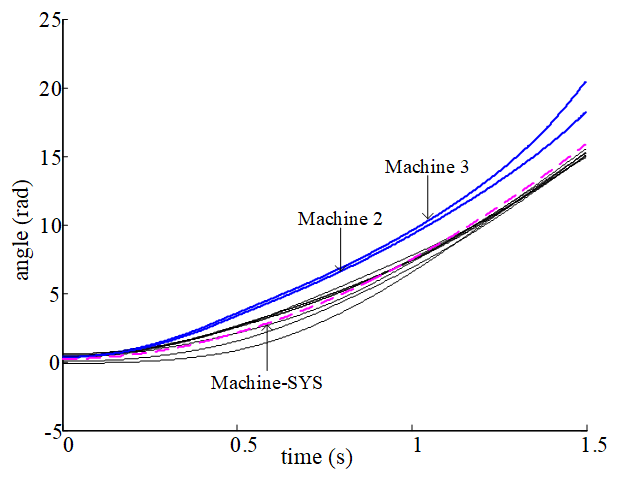}
  \caption{Original system trajectory in the synchronous reference [TS-1, bus-19, 0447 s].} 
  \label{fig11}   
\end{figure}

\begin{figure}[H]
  \centering
  \includegraphics[width=0.4\textwidth,center]{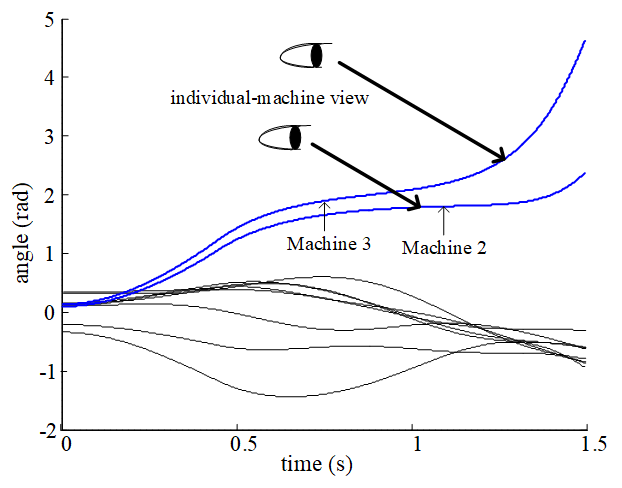}
  \caption{Original system trajectory in the COI-SYS reference  [TS-1, bus-4, 0.447 s].} 
  \label{fig12}   
\end{figure}

 Based on the analysis above, the original system trajectory is defined as the system trajectory that is formed by the IMTR of each machine in the system. A tutorial example is given below to demonstrate the mechanism of the original system trajectory.
The original system trajectory in the synchronous reference and that in the COI-SYS reference are shown in Figs. \ref{fig11} and \ref{fig12}, respectively.
\par From Fig. \ref{fig12}, the characteristics of the original system trajectory are given as below
\\\par (i) The equivalent Machine-SYS is set as the RM.
\par (ii) Using the Machine-SYS as the motion reference, the original system trajectory is formed by the IMTR of each physically real machine.
\\\par
\begin{figure}[H]
  \centering
  \includegraphics[width=0.45\textwidth,center]{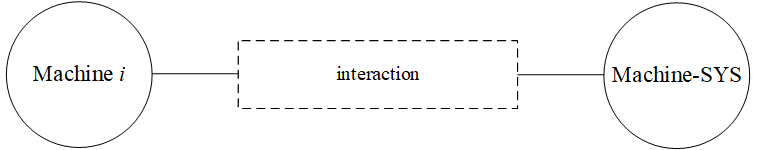}
  \caption{Formation of the CR-NCR system.} 
  \label{fig13}   
\end{figure}

 According to (i) and (ii), the IMTR of each physically real machine (in the COI-SYS reference) is still preserved in the original system trajectory. Therefore, the original system trajectory is also a multi-machine system trajectory.
\\ \textit{Individual-machine-Machine-SYS system modeling}: The variance of the $\delta_{i\mbox{-}\mathrm{SYS}}$ is modeled through the corresponding two-machine system,
i.e., the individual-machine-Machine-SYS system (I-SYS system). The formation of the I-SYS system is shown in Fig. \ref{fig13}.
Based on Eqs. (\ref{equ16}) and (\ref{equ17}), the relative motion between Machine \textit{i} and Machine-SYS is depicted as
\begin{equation}
  \left\{\begin{array}{l}
    \frac{d \delta_{i\mbox{-}\mathrm{SYS}}}{d t}=\omega_{i\mbox{-}\mathrm{SYS}} \\
    \\
    M_{i} \frac{d \omega_{i\mbox{-}\mathrm{SYS}}}{d_{t}}=f_{i\mbox{-}\mathrm{SYS}}
    \end{array}\right.
    \label{equ19} 
\end{equation}
where
\begin{spacing}{1.5}
  \noindent$f_{i\mbox{-}\mathrm{SYS}}=P_{m i}-P_{e i}-\frac{M_{i}}{M_{\mathrm{SYS}}} P_{\mathrm{SYS}}$ \\
  $\omega_{i\mbox{-}\text { SYS }}=\omega_{i}-\omega_{\text {SYS }}$
\end{spacing}
From Eq. (\ref{equ19}), the IMTR variance of the machine ($\delta_{i\mbox{-}\mathrm{SYS}}$) is completely modeled through the relative equation of motion of the I-SYS system.
\par  The individual-machine DLP (IDLP) is denoted as
\begin{equation}
  f_{i\mbox{-}\mathrm{SYS}}=0 
  \label{equ20}
\end{equation}
\par In Eq. (\ref{equ20}), the IDLP of Machine \textit{i} depicts the point where the machine becomes unstable.
\par Because the individual machine expression strictly follows the machine paradigms, The individual machine based TSA will show the following two advantages:
\\
\\Stability-characterization-advantage: The individual-machine trajectory stability is characterized precisely at MPP.
\\Trajectory-depiction-advantage: The individual-machine trajectory variance is depicted clearly at MPP.
\\\par
\begin{figure}[H]
  \centering
  \includegraphics[width=0.45\textwidth,center]{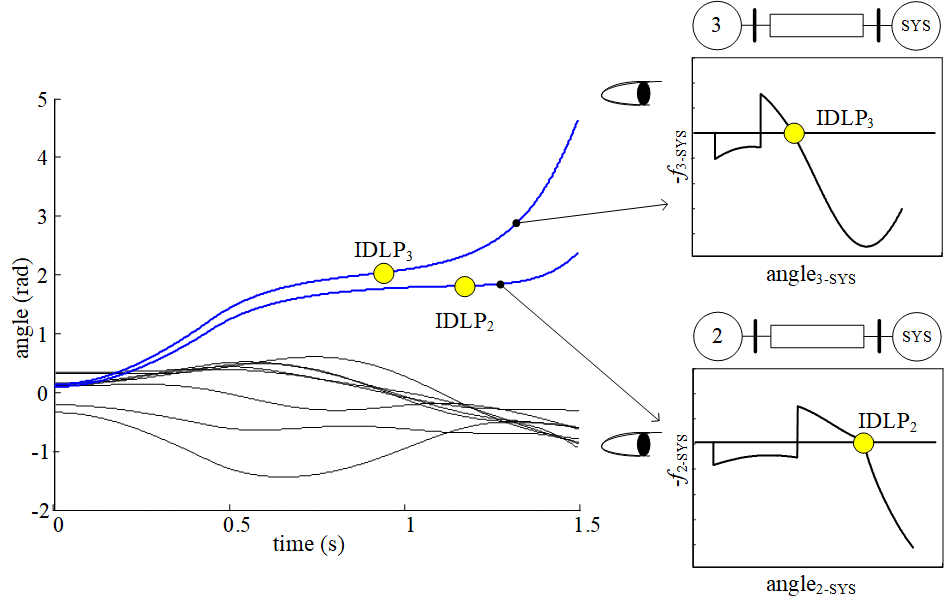}
  \caption{Individual-machine expression of the machine paradigms [TS-1, bus-4, 0.447 s].} 
  \label{fig14}   
\end{figure}

 A tutorial example is given below to demonstrate the individual-machine expression of the machine paradigms. The formations of the I-SYS system and corresponding IMEAC is shown in Fig. \ref{fig14}. From the figure, the system engineer monitors the IMTR of each machine in parallel. The stability of each machine is characterized independently through its corresponding IMEAC.

\subsection{INDIVIDUAL-MACHINE STABILITY AND ORIGINAL SYSTEM STABILITY} \label{section_VB}
\noindent \textit{Individual-machine stability}: Based on the analysis in Section \ref{section_VA}, The paradigm-based individual-machine stability characterization is given as
\\

\noindent(i) The IMTR of each machine (in the COI-SYS reference) is monitored in TSA.
\\
(ii) The variance of IMTR of each machine is modeled through its corresponding I-SYS system.
\\
(iii) The stability of each I-SYS system is evaluated through its corresponding NEC.
\\

\noindent\textit{Original System stability}: We further extend the machine stability to the system stability. For the original system with multiple machines, the stability of the entire system should be decided by the stability of each machine. Therefore, the unity principle between machine stability and system stability is depicted as below \cite{8}
\\

\noindent \textit{(Principle-I) The system can be considered to be stable if all critical machines are stable}.
\\
\textit{(Principle-II) The system can be considered to be unstable as long as any one critical machine is found to become unstable}.
\\
\par Principles I and II are defined in a genuine individual-machine manner. Principle II reveals that the entire system becoming unstable can be decided by any one critical machine becoming unstable.
\par For the multi-machine power system, the transient behavior of each machine is different, and thus the NEC inside each machine is also unique and different. Under this background, the machine will be characterized as maintaining stable or becoming unstable one after another. This fully indicates that the original multi-machine system should be evaluated in a ``machine-by-machine” manner.
\par The ``machine-by-machine” stability characterization of the original multi-machine system is given as below
\\

\noindent(i) The instability of the system is evaluated to go unstable once the ``first” IDLP (the IDLP of the leading unstable machine) occurs along time horizon.
\\ (ii) The severity of the system is obtained only when the ``last” IDLP occurs along time horizon.
\\
\par In particular, for (i), once the first DLP occurs along time horizon, the system engineer will confirm that at least one machine already becomes unstable, and thus the system is evaluated to become unstable according to the unity principle. For (ii), once the last DLP occurs along time horizon, the system engineer will finally confirm the total numbers of the unstable critical machines, and also the stability margin of each unstable critical machine. (i) and (ii) also indicate that the stability and the severity of the system will be obtained at different time points when using the individual-machine in TSA.
\par Demonstration about the ``machine-by-machine” stabiliy characterization of the original system is shown in Fig. \ref{fig13}. From the figure, the key reason for the ``machine-by-machine” action is that the transient characteristic of each machine is unique and different, as shown in Fig. \ref{fig12}.
\begin{figure}[H]
  \centering
  \includegraphics[width=0.4\textwidth,center]{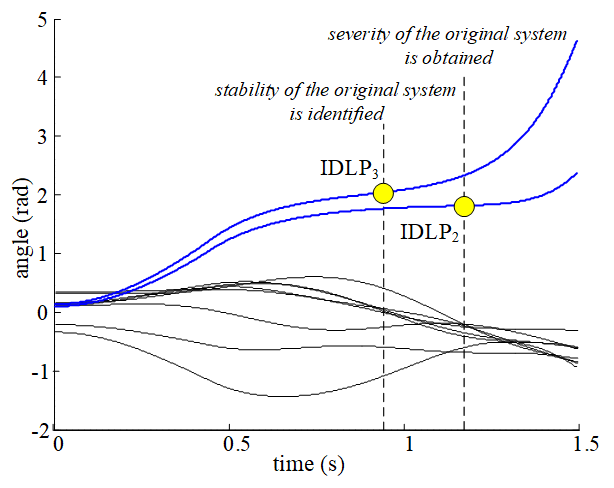}
  \caption{``Machine-by-machine” stability characterization of the original system [TS-1, bus-4, 0.447 s].} 
  \label{fig15}   
\end{figure}

\section{EQUIVALENT-MACHINE EXPRESSION} \label{section_VI}
\subsection{EQUIVALENT MACHINE TRAJECTORY} \label{section_VIA}
\noindent \textit{Equivalent system trajectory}: Motivated by the idea of the two-machine system, another formation of the system trajectory is developed. That is, the original system trajectory is first separated into two groups. Then, the two equivalent trajectories are formed in each group.
\par For a certain group separation pattern in the synchronous reference, the machine equivalence inside each group is denoted as
\begin{equation}
  \label{equ21}
  \begin{small}
  \left\{\begin{array} { l } 
    { \delta _ { \mathrm { CR } } = \frac { \sum _ { i \in \Omega _ { \mathrm { CR } } } M _ { i } \delta _ { i } } { M _ { \mathrm { CR } } } } \\
    \\
    { \omega _ { \mathrm { CR } } = \frac { \sum _ { i \in \Omega _ { \mathrm { CR } } } M _ { i } \omega _ { i } } { M _ { \mathrm { CR } } } } \\
    \\
    { M _ { \mathrm { CR } } = \sum _ { i \in \Omega _ { \mathrm { CR } } } M _ { i } }
    \end{array} \quad \left\{\begin{array}{l}
    \delta_{\mathrm{NCR}}=\frac{\sum_{j \in \Omega_{\mathrm{NCR}}} M_{j} \delta_{j}}{M_{\mathrm{NCR}}} \\
    \\
    \omega_{\mathrm{NCR}}=\frac{\sum_{j \in \Omega_{\mathrm{NCR}}} M_{j} \omega_{j}}{M_{\mathrm{NCR}}} \\
    \\
    M_{\mathrm{NCR}}=\sum_{j \in \Omega_{\mathrm{NCR}}} M_{j}
    \end{array}\right.\right.
  \end{small}
\end{equation}
\par In Eq. (\ref{equ21}), $\Omega_{\mathrm{CR}}$ and $\Omega_{\mathrm{NCR}}$ are the sets of the machines in Group-CR and Group-NCR, respectively. Theoretically, the group separation patterns are numerous. However, in this paper the group separation pattern is initially assumed as fixed for simplicity and clearance.
\par Following Eq. (\ref{equ21}), the motions of the two equivalent Machine, i.e., Machine-CR and Machine-NCR of the two groups are denoted as
\begin{equation}
  \label{equ22}
  \centering
  \left\{\begin{array}{l}
    \frac{d \delta_{\mathrm{CR}}}{d t}=\omega_{\mathrm{CR}} \\
    \\
    M_{\mathrm{CR}} \frac{d \omega_{\mathrm{CR}}}{d t}=P_{\mathrm{CR}}
    \end{array} \quad \left\{\begin{array}{l}
        \frac{d \delta_{\mathrm{NCR}}}{d t}=\omega_{\mathrm{NCR}} \\
        \\
        M_{\mathrm{NCR}} \frac{d \omega_{\mathrm{NCR}}}{d t}=P_{\mathrm{NCR}}
    \end{array} \right.  \right.
\end{equation}
where
\begin{spacing}{1.5}
  \noindent$P_{\mathrm{CR}}=\sum_{i \in \Omega_{\mathrm{CR}}}\left(P_{m i}-P_{e i}\right)$\\
  $P_{\mathrm{NCR}}=\sum_{j \in \Omega_{\mathrm{NCR}}}\left(P_{m j}-P_{e j}\right)$
\end{spacing}
From Eq.(\ref{equ22}), the EMTR variance of the machine ($\delta_{\mathrm{CR}\mbox{-}\mathrm{NCR}}$) is completely modeled through the relative equation of motion of the CR-NCR system.
The CR-NCR is formed based on the ``motion equivalence” of all machines inside each group.\\
\textit{Trajectory description}: Following the trajectory paradigm, in the COI-NCR reference, the trajectory of Machine-CR is denoted as
\begin{equation}
  \delta_{\mathrm{CR}\mbox{-}\mathrm{NCR}}=\delta_{\mathrm{CR}}-\delta_{\mathrm{NCR}}
  \label{equ23}
\end{equation}
\par In Eq. (\ref{equ23}), since Machine-NCR represents the equivalent motion of all machines in the Group-NCR, $\delta_{\mathrm{CR}\mbox{-}\mathrm{NCR}}$ going infinite along time horizon would indicate that Machine-CR becomes unstable in the COI-NCR reference.
\par
\begin{figure}[h]
  \centering
    \subfigure[]{  
      \includegraphics[width=0.4\textwidth,center]{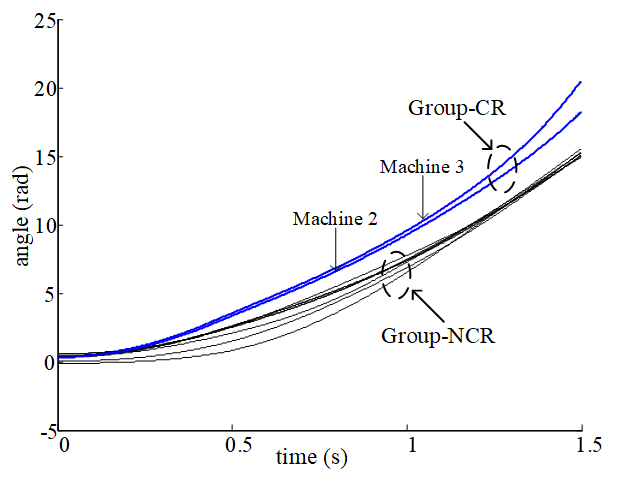}}    
      \centering
      \subfigure[]{
        \includegraphics[width=0.4\textwidth,center]{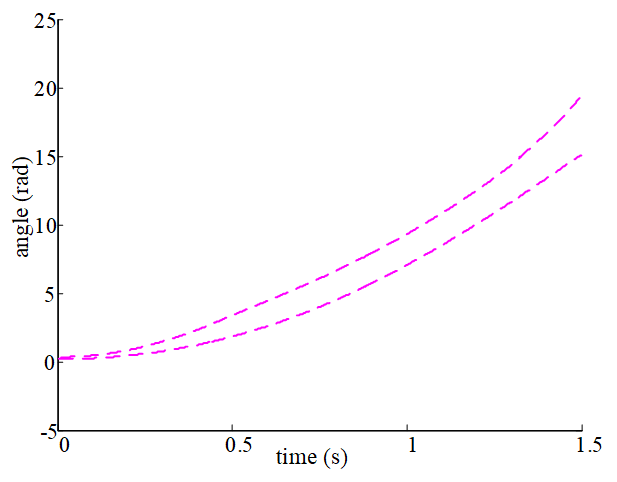}}     
   \caption{Equivalent system trajectory in the synchronous reference [TS-1, bus-4, 0.447s]. (a) Group separation. (b) Machine equivalence.}
  \label{fig16}
\end{figure}

\begin{figure}[H]
  \centering
  \includegraphics[width=0.4\textwidth,center]{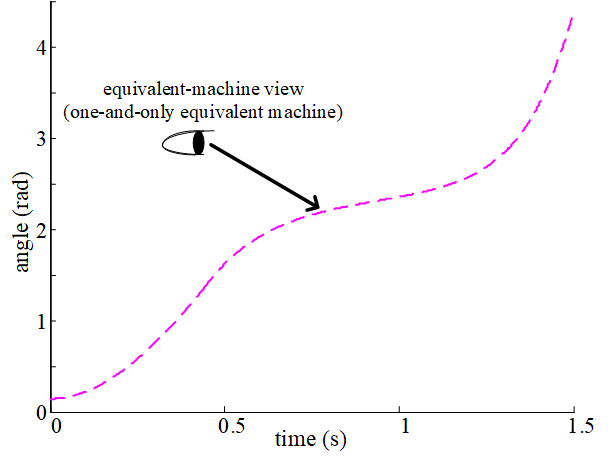}
  \caption{Equivalent system trajectory in the COI-NCR reference [TS-1, bus-4, 0.447s].} 
  \label{fig17}   
\end{figure}

 In this paper, $\delta_{\mathrm{CR}\mbox{-}\mathrm{NCR}}$ is named the equivalent-machine trajectory (EMTR). Based on this, the equivalent system trajectory is formed by the ``one-and-only” EMTR in the COI-NCR reference. A tutorial example is given below to demonstrate the equivalent system trajectory. The equivalent system trajectory in the synchronous reference and that in the COI-NCR reference are shown in Figs. \ref{fig16} and \ref{fig17}, respectively.
From Fig. \ref{fig17}, the characteristics of the equivalent system trajectory are given as below
\\
\par (i) The equivalent Machine-NCR is set as the RM.
\par (ii) Using the Machine-NCR as the motion reference, the equivalent system trajectory is formed by the ``one-and-only” EMTR of Machine-CR.
\\
\par
\begin{figure}[H]
  \centering
  \includegraphics[width=0.45\textwidth,center]{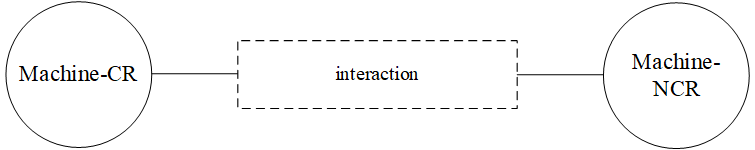}
  \caption{Formation of the CR-NCR system.} 
  \label{fig18}   
\end{figure}

 From analysis above, compared with the original system trajectory that is formed by multiple IMTRs of physically real machines, the equivalent system trajectory becomes the one-and-only $\text{EMTR}_{\text{CR}}$ in the COI-NCR reference.
\\
\textit{Machine-CR-Machine-NCR system modeling}: Following the modeling paradigm, the variance of $\delta_{\mathrm{CR}\mbox{-}\mathrm{NCR}}$ is modeled through the corresponding two-machine system, i.e., the Machine-CR-Machine-NCR system (CR-NCR system). The formation of the CR-NCR system is shown in Fig. \ref{fig18}.
Based on Eq. (\ref{equ22}), the relative motion between Machine-CR and Machine-NCR is depicted as
\begin{equation}
  \label{equ24}
  \left\{\begin{array}{l}
    \frac{d \delta_{\mathrm{CR}\mbox{-}\mathrm{NCR}}}{d t}=\omega_{\mathrm{CR}\mbox{-}\mathrm{NCR}} \\
    \\
    M_{\mathrm{CR}} \frac{d \omega_{\mathrm{CR}\mbox{-}\mathrm{NCR}}}{d t}=f_{\mathrm{CR}\mbox{-}\mathrm{NCR}}
    \end{array}\right.
\end{equation}
where
\begin{spacing}{1.5}
 \noindent $\delta_{\mathrm{CR}\mbox{-}\mathrm{NCR}}=\delta_{\mathrm{CR}}-\delta_{\mathrm{NCR}}$\\
  $\omega_{\mathrm{CR}\mbox{-}\mathrm{NCR}}=\omega_{\mathrm{CR}}-\omega_{\mathrm{NCR}}$\\
  $f_{\mathrm{CR}\mbox{-}\mathrm{NCR}}=P_{\mathrm{CR}}-\frac{M_{\mathrm{CR}}}{M_{\mathrm{NCR}}} P_{\mathrm{NCR}}$
\end{spacing}
From Eq. (\ref{equ24}), the EMTR variance of the Machine-CR ($\delta_{\mathrm{CR}\mbox{-}\mathrm{NCR}}$) is completely modeled through the relative motion in the CR-NCR system.
\par The equivalent machine DLP (EDLP) is denoted as
\begin{equation}
  \label{equ25}
  f_{\mathrm{CR}\mbox{-}\mathrm{NCR}}=0
\end{equation}
\par In Eq.(\ref{equ25}), the EDLP of Machine-CR depicts the point where the equivalent machine becomes unstable.
\par Because the equivalent machine expression strictly follows the machine paradigms, the equivalent machine based TSA will show the following two advantages:
\\
\\Stability-characterization advantage: The EMTR stability is characterized precisely at EMPP.
\\Trajectory-depiction advantage: The EMTR variance is depicted clearly at EMPP.
\\
\par
\begin{figure}[H]
  \centering
  \includegraphics[width=0.45\textwidth,center]{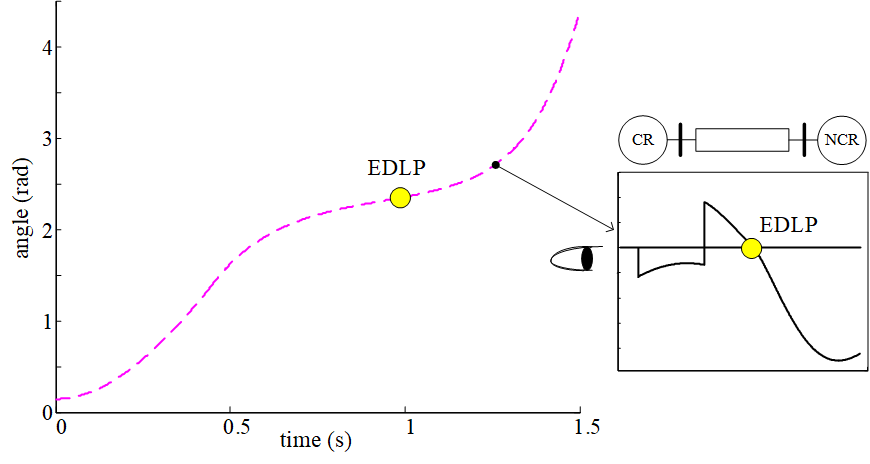}
  \caption{Equivalent-machine expression of the machine paradigms [TS-1, bus-4, 0.447 s].} 
  \label{fig19}   
\end{figure}

 A tutorial example is given below to demonstrate the equivalent-machine expression of the machine paradigms. The formation of the CR-NCR system and the corresponding EAC is shown in Fig. \ref{fig19}. From the figure, the stability of the ``one-and-only” Machine-CR is characterized through its corresponding ``one-and-only” EAC.

\subsection{EQUIVALENT MACHINE STABILITY AND EQUIVALENNT SYSTEM STABILITY} \label{section_VIB}
Based on the analysis in Section \ref{section_IVB}, the equivalent can be seen as a special ``equivalent” case of the individual machines inside the two groups. Because the equivalent machine is the ``one-and-only” machine in the equivalent system, the ``machine stability” and the ``system stability” in the equivalent system will become identical. The analysis is given as below.
\\
\textit{Equivalent-machine stability}: Based on the equivalent machine trajectory, the equivalent-machine stability evaluation is given as
\\
\\(i) The EMTR of the ``one-and-only” Machine-CR (in the COI-NCR reference) is monitored  in TSA.
\\(ii) The EMTR variance of Machine-CR is modeled through its corresponding ``one-and-only” CR-NCR system.
\\(iii) The stability of the CR-NCR system is evaluated through its corresponding ``one-and-only” NEC.
\\
\\ \textit{Equivalent system stability}: The equivalent system stability is completely identical to the ``one-and-only” equivalent machine stability.
\par Extended from the analysis in Section \ref{section_IVB}, the unity principle of the equivalent system is depicted as below
\\
\\ \textit{(Principle-I) The system can be considered to be stable if the one-and-only Machine-CR is stable}.
\\ \textit{(Principle-II) The system can be considered to be unstable if the one-and-only Machine-CR becomes unstable}.
\\
\par The two principles above can be seen as the special ``equivalent” case of the previous individual-machine based unity principle as given in Section \ref{section_IVB}.
\par Along post-fault period, since the original system trajectory is replaced with the equivalent system trajectory, the equivalent system should also be evaluated by using the ``one-and-only” Machine-CR.
\par Also based on the ``one-and-only” Machine-CR in the equivalent system, the stability characterization of the equivalent system is given as below
\\
\\ (i) The instability of the system is evaluated to go unstable once the EDLP of Machine-CR occurs along time horizon.
\\ (ii) The severity of the system is obtained once the EDLP of Machine-CR occurs along time horizon.
\\
\par
\begin{figure}[H]
  \centering
  \includegraphics[width=0.4\textwidth,center]{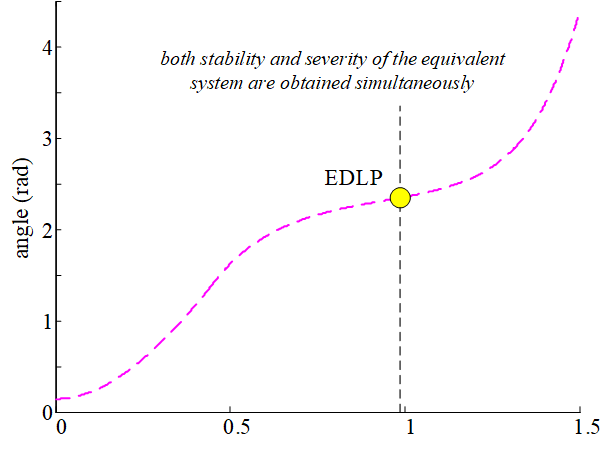}
  \caption{Stability and severity of the equivalent system [TS-1, bus-4, 0.447 s].} 
  \label{fig20}   
\end{figure}

 (i) and (ii) indicate that both the stability and the severity of the system will be obtained ``simultaneously” when using the equivalent machine in TSA.
Demonstration about the stability and severity of the original system is shown in Fig. \ref{fig20}. From the figure, the key reason for the simultaneous identification of the stability and the severity of the system is that the equivalent system is formed by the one-and-only Machine-CR. 
The equivalent-machine expression of the machine paradigms essentially originates the equivalent-machine method. A tutorial example will be given in the case study.

\section{A TUTORIAL EXAMPLE} \label{section_VII}
\subsection{INDIVIDUAL-MACHINE BASED TSA} \label{section_VIIA}
\begin{figure}[H]
  \centering
  \includegraphics[width=0.45\textwidth,center]{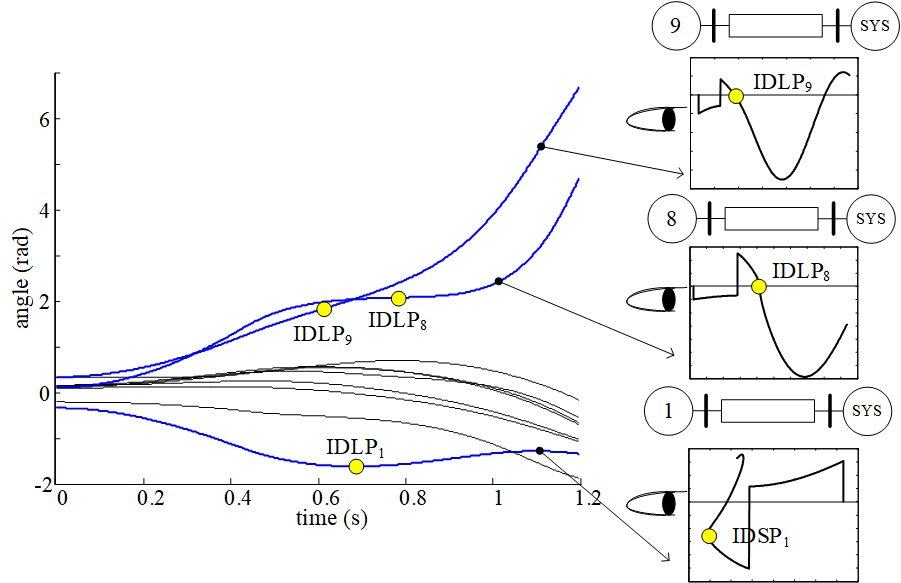}
  \caption{Original system trajectory [TS-1, bus-2, 0.430 s].} 
  \label{fig21}   
\end{figure}

The case [TS-1, bus-2, 0.430s] is provided here to demonstrate the individual-machine expression and the equivalent-machine expression in TSA. 
The original system trajectory in the COI-SYS reference is shown in Fig. \ref{fig21}. In this case Machines 8, 9 and 1 are severely disturbed critical machines.
The individual-machine expressions of the paradigms are given as below
\\
\textit{IMTR monitoring}: In the COI-SYS reference, Machines 8, 9 and 1 are severely disturbed and they are critical machines. Because critical machines are most possible to go unstable, the system engineer monitors each critical machine in parallel.
\\
\textit{I-SYS system modeling}: The IMTR of each critical machine is modeled through its corresponding I-SYS system, and thus three I-SYS systems, i.e., the $\mathrm{I\mbox{-}SYS}_{8}$, $\mathrm{I\mbox{-}SYS}_{9}$ and $\mathrm{I\mbox{-}SYS}_{1}$ are formed.
\\
\textit{IMEAC}: Based on the I-SYS system modeling of each critical machine, the IMEAC of each critical machine is formed, as in Fig. \ref{fig17}.
\\
\textit{Individual-machine stability evaluation}: The stability characterization of each critical machine is given as below
\\ $\text{IDLP}_{9}$ occurs (0.614 s): The residual $\text{IMKE}_9$ is positive (0.793 p.u.), as in Fig. \ref{fig18}. Machine 9 is evaluated as unstable. The $\text{IMTR}_{9}$ separates from the system at the moment.
\\ $\text{IDSP}_{1}$ occurs (0.686 s): Machine 1 is evaluated as stable. The $\text{IMTR}_{1}$ inflects back.
\\ $\text{IDLP}_{8}$ occurs (0.776 s): Machine 8 is evaluated as unstable. The $\text{IMTR}_{8}$ separates from the system.
\\ \textit{Original system stability evaluation}: The stability evaluation of the system is given as
\\ The stability of the system: At the moment that $\text{IDLP}_{9}$ occurs, the system is considered to become unstable according to the individual-machine based unity principle.
\\ The severity of the system: At the moment that $\text{IDLP}_{8}$ occurs, the system engineer finally confirms that Machines 9 and 8 become unstable while Machine 1 maintains stable. The severity of the entire original system is obtained as two machines in total becoming unstable.
\par  From analysis above, the individual-machine expressions of the paradigms show the two advantages in TSA. The stability of each machine is characterized precisely, and the trajectory of each machine is also depicted clearly at IMPP. 
The instability and the severity of the original system are obtained at $\text{IDLP}_{9}$ and $\text{IDLP}_{8}$, respectively. The instability of the original system is identified earlier than severity of the system. In brief, the stability of the system is evaluated in a ``machine-by-machine” manner.
The key reason for this ``machine-by-machine” stability characterization is that the transient characteristics of each individual machine is unique and different. Detailed analysis will be given in the second paper \cite{20}.

\subsection{EQUIVALENT-MACHINE BASED TSA} \label{section_VIIB}
For the case  [TS-1, bus-2, 0.430 s], the group separation pattern is set as $\Omega_{\mathrm{CR}}=\{8, 9\}$; $\Omega_{\mathrm{NCR}}=\{\mathrm{rest}\}$.
\par
\begin{figure}[H]
  \centering
  \includegraphics[width=0.43\textwidth,center]{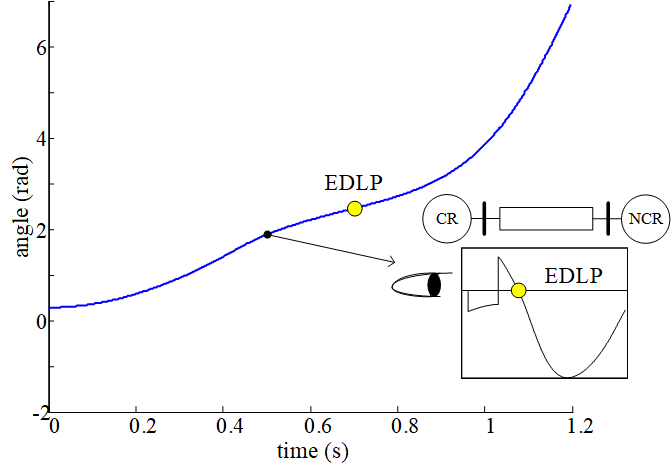}
  \caption{Equivalent system trajectory [TS-1, bus-2, 0.430 s].} 
  \label{fig22}   
\end{figure}

 The equivalent system trajectory in the COI-NCR reference is shown in Fig. \ref{fig22}.

The equivalent-machine expressions of the paradigms are given as below
\\
\textit{EMTR monitoring}: Under the given group separation pattern, Machine-CR is the one-and-only machine in the system. The system also monitors this one-and-only Machine-CR.
\\
\textit{CR-NCR system modeling}: The EMTR of Machine-CR is modeled through its corresponding one-and-only CR-NCR system.
\\
\textit{EMEAC}: Based on the CR-NCR system, the one-and-only EMEAC of Machine-CR is formed, as in Fig. \ref{fig21}.
\\
\textit{Equivalent-machine stability evaluation}: The stability characterization of each critical machine is given as below
\\
$\text{EDLP}_{\text{CR}}$ occurs (0.686 s): Machine CR is evaluated as unstable. The $\text{EMTR}_{\text{CR}}$ separates from the system.
\\ \textit{Equivalent system stability evaluation}: It is the completely the same with the equivalent machine stability evaluation.
\par From analysis above, the equivalent-machine expression of the paradigms also show advantages in TSA. In particular, the stability of each equivalent machine is characterized precisely at EMPP, and the trajectory of each equivalent machine is also depicted clearly at EMPP.
Both the instability and the severity of the equivalent system are obtained simultaneously when $\text{EDLP}_{\text{CR}}$ occurs. This is because only one equivalent Machine-CR exists in the system.
In addition, $\mathrm{EDLP}_{\mathrm{CR}}$ occurs between $\text{IDLP}_{9}$ (0.614 s) and $\text{IDLP}_{8}$ (0.776 s).
This also reflects that the equivalent system can be seen as the ``equivalence” of the original system. Detailed analysis will be given in the fourth paper \cite{22}.

\subsection{INITIAL ANALYSIS ABOUT THE INNER-GROUP STABILITY}
From the analysis in Sections \ref{section_VIIA} and \ref{section_VIIB}, since both the individual-machine and the equivalent machine are originated from the paradigms, both the stability evaluation and the trajectory variance of the machine are depicted precisely, no matter the machine is physically real or equivalent.
\par This emerges the question: what is the difference between the ``non-equivalent” individual-machine and the ``equivalent” machine?
\par
\begin{figure}[H]
  \centering
  \includegraphics[width=0.45\textwidth,center]{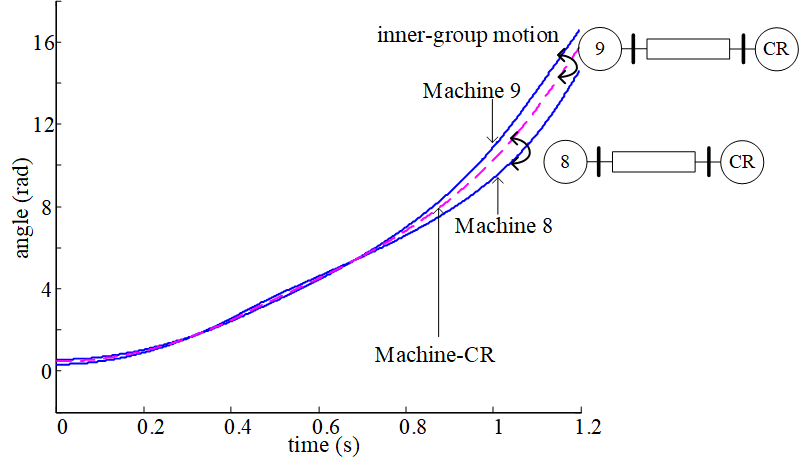}
  \caption{Inner-group motion [TS-1, bus-2, 0.430 s].} 
  \label{fig23}   
\end{figure}

 According to the primary role of the trajectory paradigm as analyzed in Section \ref{section_IVA}, it is clear that the key difference should be the difference of trajectory description. The difference between the original system trajectory and the equivalent system trajectory in the synchronous reference is shown in Fig. \ref{fig23}.
From Fig. \ref{fig23}, taking the physical machines inside Group-CR as an example, the paradigms-based inner-group TSA is given as below
\\
\\ (i) The trajectory separation inside the group can be seen as the separation between the real machine and Machine-CR.
\\ (ii) This trajectory separation can be naturally modeled through the two-machine system that is formed by the real machine and Machine-CR.
\\
\par (i) and (ii) strongly indicate that the machine paradigms can also be used in the stability characterization of the inner-group motion. In this paper, the difference between the original system trajectory and the equivalent system trajectory is named the ``inner-group trajectory”.
This trajectory belongs the ``inner-group machine”, i.e., the individual machine in the COI-CR reference. The inner-group trajectory is created from the difference between the two system trajectories. It neither exists in the original system trajectory nor be found in the equivalent system trajectory. Detailed analysis will be given in the following papers \cite{22}, \cite{23}.

\section{CONCLUSIONS} \label{section_VIII}
In this paper, the machine paradigms are established. The machine paradigms are formed by trajectory paradigm, modeling paradigm and energy paradigm in sequence. The trajectory paradigm is the reflection of the trajectory stability; the modeling paradigm is the two-machine-system modeling of the trajectory stability; and the energy paradigm is the stability evaluation of the two-machine system.
The strict followings of the paradigms will bring the two advantages in TSA. That is, the trajectory stability of the machine is characterized precisely at MPP, and the trajectory variance of the machine is also depicted clearly at MPP.
After that, the two expressions of the machine paradigms, i.e., the individual-machine expression and the equivalent-machine expression are analyzed. The two expressions are fully based on the ``non-equivalent” original system trajectory and the ``equivalent” system trajectory. The stability of the original system is characterized in a ``machine-by-machine” manner. 
Comparatively, the stability of the equivalent system is given in a ``one-and-only” machine manner. It is also clarified that the inner-group motion will be created through the difference between the original system and the equivalent system. This inner-group motion can also be analyzed through machine paradigms.
The establishment of the machine paradigms essentially explains the effectiveness of two EAC based method, i.e., the IMEAC method \cite{8} and the IEEAC method \cite{19} in TSA.
\par In the following paper, the individual machine and its strict followings of the machine paradigms will be analyzed in detail. This detailed analysis will be further used in the explanations the mechanisms of the superimposed-machine, equivalent-machine and inner-group machine who are fundamentally modeled based on the individual machine.

%

%
%
%





\begin{thebibliography}{1}
\bibitem{1}
A. A. Fouad and S. E. Stanton, ``Transient stability of a multi-multi-machine power system. Part I: Investigation of system trajectories,’’ IEEE Trans. Power App. Syst., vol. PAS-100, no. 7, pp. 3408–3416, 1981.

\bibitem{2}
A. A. Fouad and S. E. Stanton, ``Transient stability of a multi-machine power system. Part II: Critical transient energy,’’ IEEE Trans. Power App. Syst., vol. PAS-100, no. 7, pp. 3417–3424, 1981.

\bibitem{3}
V. Vittal, ``Power system transient stability using critical energy of individual machines,” Ph.D dissertation, Iowa State University, 1982.

\bibitem{4}
A. Michel, A.A. Fouad, and V. Vittal, ``Power system transient stability using individual machine energy functions,” IEEE Trans. on Circuits Syst. vol. CAS-30, no. 5, pp. 266-276, 1983.

\bibitem{5}
S. E. Stanton, ``Assessment of the stability of a multi-machine power system by the transient energy margin,” Ph.D dissertation, Iowa State University, 1982.

\bibitem{6}
S. E. Stanton and W. P. Dykas, ``Analysis of a local transient control action by partial energy functions,” IEEE Trans. Power Syst. vol. 4, no. 3, pp. 996-1002, 1989.

\bibitem{7}
S. E. Stanton, ``Transient stability monitoring for electric power systems using a partial energy function,” IEEE Trans. Power Syst. vol. 4, no. 4, pp. 1389-1396, 1989.

\bibitem{8}
S. Wang, J. Yu and W. Zhang, ``Transient stability assessment using individual machine equal area criterion PART I: unity principle”, IEEE ACCESS, vol. 6, pp. 77065-77076, 2018.

\bibitem{9}
S. Wang, J. Yu and W. Zhang, ``Transient stability assessment using individual machine equal area criterion PART II: stability margin”, IEEE ACCESS, vol. 6, pp. 38693-38705, 2018.

\bibitem{10}
S. Wang, J. Yu and W. Zhang, ``Transient stability assessment using individual machine equal area criterion PART III: reference machine”, IEEE ACCESS, vol. 7, pp. 80174-80193, 2019.

\bibitem{11}
S. Wang, J. Yu, A. Foley, and W. Zhang, ``Transient energy of an individual machine PART I: Stability Characterization”. IEEE ACCESS, vol. 9, pp. 44797-44812, 2021.

\bibitem{12}
S. Wang, J. Yu, A. Foley, and W. Zhang, ``Transient energy of an individual machine PART II: Potential Energy Surface”. vol. 9, pp. 60223-60243, 2021.

\bibitem{13}
S. Wang, J. Yu, and A. Foley, ``Transient energy of an individual machine PART III: Newtonian Energy Conversion”. vol. 9, pp. 110236-60254, 2021.

\bibitem{14}
T. Athay, R. Podmore, and S. Virmani, ``A practical method for direct analysis of transient stability,” IEEE Trans. Power Apparatus and Syst. vol. PAS-98, pp. 573-584, 1979.

\bibitem{15}
N. Kakimoto, Y. Ohsawa, and M.Hayashi, ``Transient stability analysis of electric power system via Lur’s type Lyapunov function, Part I New critical value for transient stability,” IEE Trans. Japan, Vol. 98, no. 5/6, pp. 63-71, 1978.

\bibitem{16}
N. Kakimoto, Y. Ohsawa, and M.Hayashi, ``Transient stability analysis of electric power system via Lur’s type Lyapunov function, Part II Modification of Lure Type Liapunov Function with Effect of Transfer Conductances,” IEE Trans. Japan, Vol. 98, no. 5/6, pp. 72-79, 1978.

\bibitem{17}
Y. Xue, ``Re-examination of transient energy functions and critical energy”, Automation of electric power systems, vol. 6, pp. 9-18, 1991.

\bibitem{18}
D. Fang, T. S. Chung, Y. Zhang, and W. Song, ``Transient stability limit conditions analysis using a corrected transient energy function approach”, IEEE Trans. Power Syst. vol. 15, no. 2, pp. 804-810, 2000.

\bibitem{19}
Y. Xue, ``Integrated extended equal area criterion–Theory and application,” in Proc. 5th Symp. Specialists in Electric Operational and Expansion Planning, Recife, Brazil, 1996.

\bibitem{20}
S. Wang, J. Yu, and A. Foley, ``Newtonian Mechanics Based Transient Stability PART II: Individual Machine”.

\bibitem{21}
S. Wang, J. Yu, and A. Foley, ``Newtonian Mechanics Based Transient Stability PART III: Superimposed Machine”.

\bibitem{22}
S. Wang, J. Yu, and A. Foley, ``Newtonian Mechanics Based Transient Stability PART IV: Equivalent Machine”.

\bibitem{23} 
S. Wang, J. Yu, and A. Foley, ``Newtonian Mechanics Based Transient Stability PART V: Inner-group Machine”.

\bibitem{24}
S. Wang, J. Yu, and A. Foley, ``Newtonian Mechanics Based Transient Stability PART VI: Machine Transformation”.

\end{thebibliography}
\end{document}